\documentclass[11pt]{article}

\usepackage[T1]{fontenc}
\usepackage[centertags]{amsmath}
\usepackage{amsfonts}
\usepackage{amssymb}
\usepackage[pdftex]{hyperref}
\usepackage{graphicx}
\usepackage{graphbox}
\usepackage[numbers,sort&compress]{natbib}

\usepackage{paperinitial}

\paperinitialization{15mm}{15mm}{15mm}{15mm}{2pt}{10pt}

\DeclareMathOperator{\im}{Im}
\DeclareMathOperator{\sgn}{sgn}
\DeclareMathOperator{\tg}{tg}
\DeclareMathOperator{\ctg}{ctg}
\DeclareMathOperator{\pr}{pr}

\DeclareMathOperator{\Texp}{Texp}

\newcommand{\lan}{\langle}
\newcommand{\ran}{\rangle}

\newcommand{\e}{\varepsilon}
\newcommand{\vf}{\varphi}
\newcommand{\vk}{\varkappa}
\newcommand{\s}{\sigma}
\newcommand{\Si}{\Sigma}
\newcommand{\al}{\alpha}
\newcommand{\be}{\beta}
\newcommand{\ga}{\gamma}

\newcommand{\de}{\delta}
\newcommand{\De}{\Delta}

\newcommand{\la}{\lambda}

\newcommand{\spx}{\mathbf{x}}
\newcommand{\spy}{\mathbf{y}}

\newcommand{\spe}{\mathbf{e}}
\newcommand{\spb}{\mathbf{b}}

\begin{document}
\allowdisplaybreaks[4]
\frenchspacing
\setlength{\unitlength}{1pt}

\title{{\Large\textbf{Probability of radiation of twisted photons in an inhomogeneous isotropic dispersive medium}}}

\date{}

\author{O.V. Bogdanov${}^{1),2)}$\thanks{E-mail: \texttt{bov@tpu.ru}},\; P.O. Kazinski${}^{1)}$\thanks{E-mail: \texttt{kpo@phys.tsu.ru}},\; and G.Yu. Lazarenko${}^{1)}$\thanks{E-mail: \texttt{laz@phys.tsu.ru}}\\[0.5em]
{\normalsize ${}^{1)}$ Physics Faculty, Tomsk State University, Tomsk 634050, Russia}\\
{\normalsize ${}^{2)}$ Tomsk Polytechnic University, Tomsk 634050, Russia}}

\maketitle

\begin{abstract}

The general formula for probability to record a twisted photon produced by a charged particle moving in an inhomogeneous isotropic dispersive medium is derived. The explicit formulas for probability to record a twisted photon are obtained for the radiation of a charged particle traversing a dielectric plate or an ideally conducting foil. It is shown that, in the case when the charged particle moves along the detector axis, all the radiated twisted photons possess a zero projection of the total angular momentum and the probability of their radiation is independent of the photon helicity. The radiation produced by helically microbunched beams of charged particles is also considered. The fulfillment of the strong addition rule for the projection of the total angular momentum of radiated twisted photons is demonstrated. Thus the helical beams allow one to generate coherent transition and Vavilov-Cherenkov radiations with large projections of the total angular momentum. The radiation produced by charged particles in a helical medium is studied. Typical examples of such a medium are metallic spirals and cholesteric liquid crystals. It is shown that the radiation of a charged particle moving along the helical axis of such a medium is a pure source of twisted photons.

\end{abstract}

\section{Introduction}

The use of media with nontrivial permittivity is the most common method to generate twisted photons \cite{KnyzSerb,PadgOAM25,Roadmap16,AndBabAML,TorTorTw,AndrewsSLIA,VoloLavr,MarManPap,Baboza12,LouDelBra13,Barboza13,Migara18,KraPodKacBra,YLiu18}. However, as a rule, the medium is employed only as a converter of plane-wave photons to twisted ones. We investigate in the present paper a direct means for production of twisted photons by charged particles moving in an inhomogeneous dispersive medium. As far as a homogeneous medium is concerned, the theory of radiation of twisted photons is known in this case (see the description of the Vavilov-Cherenkov process in \cite{IvSerZay,Kaminer16}). However, the detectors of twisted photons \cite{LPBFAC,BLCBP,SSDFGCY,LavCourPad,RGMMSCFR,PSTP19} or the objects that should be irradiated by them \cite{MHSSF13,PesFriSur15,AfSeSol18,SGACSSK} are usually positioned outside of the medium. When the twisted photons escape from the medium, the form of their peculiar phase front can be destructed. Therefore, to describe properly the production of twisted photons by charged particles moving in inhomogeneous media, the corresponding theory has to be constructed.

The classical and quantum theories of radiation by particles propagating in dispersive media are well elaborated (see, e.g., \cite{Cherenkov,Vavilov,TammFrank37,Ginz40,Sokol40,SokolLosk,SokolLosk1,Riazanov57,Riazanov58,AbrGorDzyal,AleksNik,TMikaelian,SchwinTsai,GinzbThPhAstr,Frank84,LandLifshECM,RyazanovB,BazylZhev,GinzTsyt,GlaubLewen,Orisa94,AfanasBo,Khripl09,Tuchin18}). So we adapt these theories for description of radiation of twisted photons. First of all, we develop quantum electrodynamics (QED) in an inhomogeneous dispersive medium with twisted photons and derive the general formula for the probability to record a twisted photon created by a classical current. The range of applicability of such a theory is, of course, the same as for the corresponding theory of radiation of plane-wave photons. Then, employing this formalism, we investigate several representative examples \cite{AfanasBo,TMikaelian,GinzTsyt,Pafomov,BarysFran,GinzbThPhAstr} to demonstrate the main features of twisted photon radiation by charged particles in a medium. Namely, we consider the radiation of twisted photons by a charged particle moving uniformly along a straight line intersecting a dielectric plate or a metal foil. With the aid of the results of \cite{BKb,BKLb}, we investigate the coherence of such a radiation created by particle beams and the conditions when transition and Vavilov-Cherenkov (VCh) radiations can be used as a pure bright source of twisted photons. In particular, we show that the helically microbunched beams \cite{RibGauNin14,HemMar12,HemMarRos11,HemRos09,HemStuXiZh14,HKDXMHR,HemsingTR12,CLiu16,XLZhu18,LBJu16} with a certain helix pitch and transverse size produce coherent twisted transition and VCh radiations with large projection of the total angular momentum. Notice that coherent transition and VCh radiations of plane-wave photons were discussed in many papers and books (see, e.g., \cite{BazylZhev,ArutOgan94,KuzRukh08,WenXin11,GevIspSham,AlTyuGal19,ACurcio19}). As for helically microbunched beams, they were already used to generate twisted photons by undulators \cite{HemMarRos11,HemMar12,HKDXMHR,RibGauNin14,HemStuXiZh14} and by hitting metal foils \cite{HemsingTR12,HemRos09}. The peculiarities of twisted photon radiation produced by particle bunches of different profiles can also be employed for diagnostics of the particle beam structure \cite{HemsingTR12,HemRos09,HLarocque16}.

Another pure source of twisted photons that we study in the present paper is a helical medium. Typical examples of such a medium are metallic helical ribbons \cite{YLiu18}, cholesteric liquid crystals \cite{Barboza15,Baboza12,VoloLavr}, and helically arranged dielectrics. Usually, the helical media are used to convert ordinary plane-wave photons to twisted ones. We, however, consider the direct process of radiation by charged particles moving in such a medium. It turns out that a charged particle moving uniformly along the helix axis of such a medium is a pure source of twisted photons. Its radiation obeys the selection rules that are pertinent to ideal helical undulators \cite{Rubic17,KatohPRL,TaHaKa,SasMcNu,TairKato,BKL4,BKL2,BKL3,BHKMSS,BKb,Hemsing14hel,KatohSRexp} or to scattering on helical targets \cite{BKL3}. This fact has a simple explanation in terms of transition scattering of the permittivity wave on the charged particle. This source can be employed to generate twisted photons with energies up to X-ray spectral range. Of course, there are other pure sources of twisted photons such as undulators and laser waves where the twisted photons are produced directly by charged particles \cite{BKL4,EppGusel19,AfanMikh,BordKN,TaHaKa,KatohPRL,TairKato,BKL2,KatohSRexp,Rubic17,Hemsing14hel,BHKMSS,SasMcNu,HemStuXiZh14,RibGauNin14,HKDXMHR,HemMar12,HemMarRos11}. The hard twisted photons with energies of order of hundreds MeVs can be generated by inverse Compton scattering of low energy twisted photons \cite{JenSerprl,JenSerepj} or in channeling \cite{ABKT,EpJaZo}.

The paper is organized as follows. In Sec. \ref{Quant_in_Medium}, we develop QED in an inhomogeneous dispersive medium. In Sec. \ref{Recording_Tw_Phot}, we find the probability to record a twisted photon by a detector in a vacuum and obtain the general formula for the radiation probability of twisted photons by classical currents. In Sec. \ref{Homogeneous_Medium}, we obtain the general formula for the probability to record a twisted photon created by a charged particle moving in a homogeneous medium and reproduce the known results for VCh radiation of twisted photons in such media. In Sec. \ref{Examples}, we apply the general theory to the particular examples. In Secs. \ref{Diel_Plate}, \ref{Thick_Diel_Plate}, we consider the radiation of twisted photons by a charged particle or a beam of them that traverses a dielectric plate. In Sec. \ref{Cond_Plate}, transition radiation by a charged particle hitting a metal foil is investigated. The radiation produced by charged particles in helical media is studied in Sec. \ref{Hel_Medium}. The evaluation of the incoherent and coherent interference factors \cite{BKb,BKLb} for Gaussian helically microbunched beams is removed to Appendix \ref{Inter_Fact_App}. In conclusion section, we summarize the results.

We work in the system of units such that $\hbar=c=1$ and $e^2=4\pi\al$, where $\al\approx1/137$ is the fine structure constant. The notation from \cite{BKL2} is vastly employed.

\section{Quantum electromagnetic field in a medium}\label{Quant_in_Medium}

For the reader convenience and concordance of notation, in this section we develop QED in an inhomogeneous transparent isotropic dispersive medium. As regards the homogeneous medium, such a construction is well known (see, e.g., \cite{Ginz40,Sokol40,SokolLosk,SokolLosk1,Riazanov57,Riazanov58,AbrGorDzyal,AleksNik,SchwinTsai,GinzbThPhAstr,LandLifshECM,RyazanovB,GinzTsyt,Orisa94,Tuchin18}). We suppose that the spatial dispersion is negligible, the magnetic permeability $\mu=1$, and the permittivity $\e(k_0,\spx)>0$. In particular, $\im \e(k_0,\spx)=0$, where $k_0\in \mathbb{R}$ is found from solution of the Maxwell equations \eqref{Max_eqns}, i.e., it is ``on-shell''. The absence of absorption is a necessary requirement for quantum field theory to be unitary. The generalization to the case of a medium with small absorption will be given below. The procedure developed below is analogous to the one used in \cite{BKL2} for description of radiation of twisted photons by classical currents in a vacuum.

The free Maxwell equations in a medium have the form (see, e.g. \cite{LandLifshECM,RyazanovB,AbrGorDzyal})
\begin{equation}\label{Max_eqns}
    (k_0^2\e(k_0,\spx)-\hat{h}^2)A_i(k_0,\spx)=0,\qquad \partial_i(\e(k_0,\spx) A_i(k_0,\spx))=0,
\end{equation}
where $A_i(k_0,\spx)$ is the Fourier transform of the vector potential, and we have introduced the Maxwell Hamiltonian (the curl operator)
\begin{equation}
    \hat{h}_{ij}:=\e_{ikj}\partial_k.
\end{equation}
The second condition in \eqref{Max_eqns} is the generalization of the Coulomb gauge. It ensues from the first equation in \eqref{Max_eqns} provided $k_0\neq0$. Further, we assume that the electromagnetic field obeys such boundary conditions that $k_0>0$. Besides, we suppose that $\e(k_0,\spx)\neq1$ in the region $M$ of a finite volume. The rest part of space possessing an infinite volume will be denoted as $\Omega$. Since we assume $\im \e(k_0,\spx)=0$ on-shell then, for these values of $k_0$ \cite{LandLifshECM}
\begin{equation}
    \e(k_0,\spx)=\e(-k_0,\spx).
\end{equation}
Introducing the operator
\begin{equation}\label{Max_sqrd}
    \hat{\tilde{h}}^2(k_0):=\e^{-1/2}\hat{h}^2\e^{-1/2},
\end{equation}
we cast equations \eqref{Max_eqns} into an explicitly self-adjoint form
\begin{equation}\label{Max_eqns1}
    (k_0^2-\hat{\tilde{h}}^2)\tilde{A}_i(k_0,\spx)=0,\qquad \partial_i(\e^{1/2}(k_0,\spx) \tilde{A}_i(k_0,\spx))=0,
\end{equation}
where $\tilde{A}_i(k_0,\spx)=\e^{1/2}(k_0,\spx) A_i(k_0,\spx)$. The operators entering into equations \eqref{Max_eqns1} remain unchanged under the replacement $k_0\rightarrow-k_0$.

It is not difficult to find the general solution of equations \eqref{Max_eqns1} for real $k_0$ such that $\im \e(k_0,\spx)=0$. Let us pose the eigenvalue problem
\begin{equation}\label{eigen_probl}
    \hat{\tilde{h}}^2(k_0)\tilde{\psi}_\al(k_0)=\chi^2_\al(k_0)\tilde{\psi}_\al(k_0),\qquad \chi_\al(k_0)\geqslant0,
\end{equation}
where $\al$ marks the eigenvalues $\chi^2_\al(k_0)$. The operator on the left-hand side of \eqref{eigen_probl} is self-adjoint in the Hilbert space of complex vectors $\tilde{\psi}_i(k_0;\spx)$ subject to the condition
\begin{equation}
    \partial_i(\e^{1/2}(k_0)\tilde{\psi}_i(k_0))=0,
\end{equation}
with the standard scalar product
\begin{equation}\label{scal_prod}
    \lan\tilde{\vf},\tilde{\psi}\ran=\int d\spx\tilde{\vf}^*_i(\spx)\tilde{\psi}_i(\spx).
\end{equation}
The eigenfunctions \eqref{eigen_probl} constitute a complete orthonormal set. The completeness relation reads
\begin{equation}\label{compl_rels}
    \sum_\al \tilde{\psi}_{\al i}(k_0;\spx) \tilde{\psi}^*_{\al j}(k_0;\spy)=\de^\perp_{ij}(k_0;\spx,\spy)= \big\{ \de_{ij}-\e^{1/2}(k_0,\spx)\partial_i \De_\e^{-1} \partial_j\e^{1/2}(k_0,\spy) \big\} \de(\spx-\spy),
\end{equation}
where $\De_\e^{-1}$ is the inverse to the operator $\De_\e:=\partial_i\e(k_0,\spx)\partial_i$. Taking complex conjugate of \eqref{eigen_probl}, wee see that the complete set contains the function $\tilde{\psi}^*_\al$ along with $\tilde{\psi}_\al$ corresponding to the same eigenvalue $\chi^2_\al$ but with the different quantum number $\al'(\al)$. In other words,
\begin{equation}\label{mode_conj}
    \tilde{\psi}^*_\al(k_0)=\tilde{\psi}_{\al'(\al)}(k_0),\qquad \chi_\al(k_0)=\chi_{\al'(\al)}(k_0),\qquad \al'(\al'(\al))=\al,
\end{equation}
and $\al'(\al)=\al$ for real-valued $\tilde{\psi}_{\al}(k_0)$. In particular, it follows from this property that the left-hand side of \eqref{compl_rels} is real.

If the permittivity is discontinuous on some closed hypersurface $\Si$, then the standard boundary conditions for the electromagnetic field strength on $\Si$ turn into
\begin{equation}\label{bound_conds}
    [\psi_\tau]=0,\qquad[(\hat{h}\psi)_\tau]=0,
\end{equation}
where the square brackets denote a discontinuity jump of the corresponding function on the hypersurface $\Si$ and the index $\tau$ means that only tangent to $\Si$ components of the complex vector $\psi$ should be taken. It is implied in \eqref{bound_conds} that the surface charge and the current density are absent on $\Si$. The conditions \eqref{bound_conds} entail, in particular, that \cite{AbrGorDzyal}
\begin{equation}
    [\hat{h}\psi]=0,\qquad[\e\psi_n]=0,
\end{equation}
where $n$ denotes the vector component normal to $\Si$. The boundary conditions \eqref{bound_conds} are consistent with equations \eqref{Max_eqns1} in a sense that they make the operator $\hat{\tilde{h}}^2$ self-adjoint in the space of divergence-free complex vector fields obeying \eqref{bound_conds}. The corresponding boundary terms (the singular current) for the fields defined in the region bounded by the surface $\Si$ have the form \eqref{sing_sourc}. The fields defined from the outside of $\Si$ result in the same singular current but with an opposite sign ($n^i\rightarrow-n^i$). The fulfillment of boundary conditions \eqref{bound_conds} leads to cancelation of the singular currents. Notice that the boundary conditions on the surface of an ideal conductor look as
\begin{equation}\label{bound_conds_cond}
    \psi_\tau=0,
\end{equation}
and $\psi=0$ inside of the conductor. The operator $\hat{\tilde{h}}^2$ is self-adjoint on account of these boundary conditions.

Assuming that operator \eqref{Max_sqrd} does not have zero eigenvalues, the complete set \eqref{eigen_probl} allows one to write the quantum field satisfying \eqref{Max_eqns1}:
\begin{equation}\label{quant_fields}
    \hat{\tilde{A}}_i(t,\spx)=\sum_\al\Big[\hat{a}_\al f^{1/2}_\al \tilde{\psi}_{\al i}(k_{0\al};\spx) e^{-ik_{0\al}t} +\hat{a}^\dag_\al f^{1/2}_\al \tilde{\psi}^*_{\al i}(k_{0\al};\spx)e^{ik_{0\al}t} \Big].
\end{equation}
Here
\begin{equation}\label{CAO}
    [\hat{a}_\al,\hat{a}_\be]=[\hat{a}^\dag_\al,\hat{a}^\dag_\be]=0,\qquad [\hat{a}_\al,\hat{a}^\dag_\be]=\de_{\al\be},
\end{equation}
the on-shell condition takes the form
\begin{equation}\label{mass_shell}
    k^2_{0\al}=\chi^2_\al(k_{0\al}),\qquad k_{0\al}>0,
\end{equation}
and the normalization coefficients $f_\al>0$ are found from the requirement that the residue of the propagator of the field \eqref{quant_fields} is equal to unity. The quantum field $\hat{A}_i(x)$ is obtained from \eqref{quant_fields} in an obvious manner
\begin{equation}\label{em_qfield}
    \hat{A}_i(t,\spx)=\sum_\al\Big[\hat{a}_\al f^{1/2}_\al \psi_{\al i}(k_{0\al};\spx) e^{-ik_{0\al}t} +\hat{a}^\dag_\al f^{1/2}_\al \psi^*_{\al i}(k_{0\al};\spx)e^{ik_{0\al}t} \Big],
\end{equation}
where $\psi_{\al i}(k_{0\al};\spx)=\e^{-1/2}(k_{0\al};\spx) \tilde{\psi}_{\al i}(k_{0\al};\spx)$.

The commutator Green function is written as
\begin{equation}
    \tilde{G}_{ij}(t,\spx;t',\spy):=[\hat{\tilde{A}}_i(t,\spx),\hat{\tilde{A}}_j(t',\spy)]=\sum_\al f_\al\Big[ \tilde{\psi}_{\al i}(k_{0\al};\spx) \tilde{\psi}^*_{\al j}(k_{0\al};\spy) e^{-ik_{0\al}(t-t')} -c.c. \Big].
\end{equation}
On the other hand, the retarded Green function for the operator \eqref{Max_eqns1} is
\begin{equation}\label{ret_Gr_func}
    G^{-}_{ij}(t,\spx;t',\spy)=\int\frac{d k_0}{2\pi} \sum_\al \tilde{\psi}_{\al i}(k_0;\spx)\frac{e^{-i k_0 (t-t')}}{k_0^2-\chi^2_\al(k_0)} \tilde{\psi}^*_{\al j}(k_0;\spy),
\end{equation}
where the integration contour over $k_0$ runs a little bit higher than the real axis. Taking into account that
\begin{equation}
    G^{-}_{ij}(t,\spx;t',\spy)=-i\theta(t-t')\tilde{G}_{ij}(t,\spx;t',\spy),
\end{equation}
and evaluating \eqref{ret_Gr_func} for $t>t'$ by residues, we obtain \cite{GinzbThPhAstr,LandLifshECM,GinzTsyt,MigdalMM,FursVass,KalKaz3}
\begin{equation}\label{norm_fctr}
    f^{-1}_\al=[k_0^2-\chi^2_\al(k_0)]'_{k_0=k_{0\al}}=2k_{0\al}[1-\chi'_\al(k_{0\al})],
\end{equation}
where $k_{0\al}$ is the solution of \eqref{mass_shell}. For the quantum theory to be unitary, the right-hand side of \eqref{norm_fctr} must be positive for $k_{0\al}>0$. Notice that this condition is fulfilled for a homogeneous isotropic medium due to the inequalities that hold for the permittivity $\e(k_0)$ and its derivative  (see \cite{LandLifshECM}, Sec. 84, and Sec. \ref{Homogeneous_Medium} below).

Denote as
\begin{equation}
    \bar{\de}_{ij}(\spx,\spy):=\sum_\al \tilde{\psi}_{\al i}(k_{0\al};\spx) \tilde{\psi}^*_{\al j}(k_{0\al};\spy).
\end{equation}
As long as $k_{0\al}=k_{0\al'(\al)}$ and relations \eqref{mode_conj} hold, we have
\begin{equation}
    \bar{\de}^*_{ij}(\spx,\spy)=\bar{\de}_{ij}(\spx,\spy).
\end{equation}
The operator $\bar{\de}$ is identity on the space of solutions to the Maxwell equations \eqref{Max_eqns1} in a sense that
\begin{equation}
    \int d\spy\bar{\de}_{ij}(\spx,\spy)\hat{A}_j(t,\spy)=\hat{A}_i(t,\spx).
\end{equation}
Furthermore, it follows from \eqref{mode_conj} that
\begin{equation}
    \tilde{G}_{ij}(t,\spx;t,\spy)=[\hat{\tilde{A}}_i(t,\spx),\hat{\tilde{A}}_j(t,\spy)]=0.
\end{equation}
By analogy with quantum field theory in a vacuum, one can introduce the canonical momentum
\begin{equation}
    \hat{\tilde{\pi}}_i(t,\spx):=-\frac{i}{2}\sum_\al \Big[\hat{a}_\al f^{-1/2}_\al \tilde{\psi}_{\al i}(k_{0\al};\spx) e^{-ik_{0\al}t} -\hat{a}^\dag_\al f^{-1/2}_\al \tilde{\psi}^*_{\al i}(k_{0\al};\spx)e^{ik_{0\al}t} \Big].
\end{equation}
Then
\begin{equation}
    [\hat{\tilde{A}}_i(t,\spx),\hat{\tilde{\pi}}_j(t,\spy)]=i\bar{\de}_{ij}(\spx,\spy), \qquad[\hat{\tilde{\pi}}_i(t,\spx),\hat{\tilde{\pi}}_j(t,\spy)]=0.
\end{equation}
The last property follows from relations \eqref{mode_conj}. The evolution operator of the electromagnetic field in the presence of the external conserved classical current is given by (see, e.g., \cite{WeinbergB.12})
\begin{equation}
    \hat{U}_{t_2,t_1}=\hat{U}^0_{t_2,0}\hat{S}_{t_2,t_1}\hat{U}^0_{0,t_1},\qquad \hat{S}_{t_2,t_1}=\Texp\Big[-i\int_{t_1}^{t_2}dx\hat{A}_i(x) j^i(x)-i\int_{t_1}^{t_2}dtV_{\text{Coul}}\Big],
\end{equation}
where $\hat{U}^0_{t_2,t_1}$ is the evolution operator of a free electromagnetic field \eqref{em_qfield} and $V_{\text{Coul}}$ is the energy of Coulomb self-interaction of the current. In the case at hand, this energy gives only the contribution to the common phase of transition amplitudes, and we do not take it into account anymore. It is noteworthy, however, that this phase is divergent in the infrared limit, if the system possesses an uncompensated charge \cite{WeinbIR,WeinbergB.12}.

\section{Recording twisted photons}\label{Recording_Tw_Phot}

Having constructed quantum field theory in a medium, we ought to define what we mean under the twisted photon and its recording by the detector. Suppose that the detector of twisted photons is located in the region $\Omega$, which has an infinite volume, and $\e(k_0,\spx)=1$ in it. Let $\tilde{\psi}_\al(k_{0\al};\spx)$ be a complete set of solutions to the Maxwell equations described in the previous section. Using these mode functions, we construct the wave packets $\vf_\be$ such that
\begin{enumerate}
  \item $\vf_\be\approx\vf_{0\be}$ in the vicinity of the detector of twisted photons, where $\vf_{0\be}$ are the modes corresponding to twisted photons in a vacuum (see the explicit expression in \cite{JenSerprl,JenSerepj,Ivanov11,GottfYan,JaurHac,BiaBirBiaBir,BKL2} and \eqref{tw_phot_vac});
  \item $\vf_\be$ are sufficiently narrow with respect to the energy quantum number, i.e., they are composed from the mode functions $\tilde{\psi}_\al(k_{0\al};\spx)$ with small dispersion of energy $\De k_{0\al}$.
\end{enumerate}
It is the photon state $\vf_\be$ which is assumed to be recorded by the detector. These conditions imply, in particular, that the detector is positioned in the wave zone and the region, where $\vf_\be\approx\vf_{0\be}$, is sufficiently large. As follows from the uncertainty relation,
\begin{equation}
    \De k_{0\al}\sim 2\pi/L_{vac},
\end{equation}
where $L_{vac}$ is a typical size of this region.

In the vicinity of the detector, under the above conditions, we can write
\begin{equation}\label{field_op1}
    \hat{A}_i(t,\spx)\approx\int d\spy\sum_\be \vf_{\be i}(\spx)\vf^*_{\be j}(\spy)\hat{A}_j(t,\spy)=\sum_{\al,\be}\big[\vf_{\be i}(\spx) f_\al^{1/2}\hat{a}_\al \lan\vf_\be,\psi_\al(k_{0\al})\ran e^{-ik_{0\al}t} +h.c.\big].
\end{equation}
Assuming that
\begin{equation}
    \De k_{0\be} T\ll1,
\end{equation}
we obtain
\begin{equation}\label{field_op2}
    \hat{A}_i(t,\spx)\approx\sum_{\al,\be}\big[\vf_{\be i}(\spx)e^{-ik_{0\be}t} f_\al^{1/2}\hat{a}_\al \lan\vf_\be,\psi_\al(k_{0\al})\ran  +h.c.\big].
\end{equation}
The parameter $T$ is the observation period. In fact, in order to describe the radiation correctly, we need the fulfillment of a weaker condition in \eqref{field_op1}, \eqref{field_op2}:
\begin{equation}
    \De k_{0\be} t_{f}\ll1,
\end{equation}
where $t_f$ is the radiation formation time. The fulfillment of the last condition can always be achieved by using the detector of twisted photons with sufficiently narrow bandwidth. Introduce the notation
\begin{equation}
    \hat{b}_\be:=\sqrt{2k_{0\be}}\sum_{\al} f_\al^{1/2}\hat{a}_\al \lan\vf_\be,\psi_\al(k_{0\al})\ran.
\end{equation}
Then, in the vicinity of the detector, the quantum field becomes
\begin{equation}
    \hat{A}_i(t,\spx)=\sum_{\be}\big[\vf_{\be i}(\spx)\frac{e^{-ik_{0\be}t}}{\sqrt{2k_{0\be}}}\hat{b}_\be+h.c.\big],
\end{equation}
which coincides with the decomposition of the free quantum electromagnetic field in terms of twisted photons.

In the first Born approximation with respect to the external current, the transition amplitude of the process
\begin{equation}\label{0g}
    0\rightarrow \ga
\end{equation}
reads as
\begin{equation}\label{amplitude1}
    S(\be;0)=\lan0|\hat{b}_\be \hat{U}_{T/2,-T/2}|0\ran\approx-i e^{-iT E_{vac}}\sqrt{2k_{0\be}}\sum_\al\int_{-T/2}^{T/2} dy f_\al  \lan\vf_\be,\psi_\al(k_{0\al})\ran \psi^*_{\al i}(\spy)j^i(y)e^{-ik_{0\al}(T/2-y^0)},
\end{equation}
where it is assumed that $\hat{H}_0|0\ran=E_{vac}|0\ran$. Let us introduce the positive-frequency Green function
\begin{equation}
    G^{(+)}_{ij}(x,y):=-i\lan0|\hat{A}_i(x)\hat{A}_j(y)|0\ran= -i\sum_\al f_\al e^{-ik_{0\al}(x^0-y^0)}\psi_{\al i}(k_{0\al};\spx) \psi^*_{\al j}(k_{0\al};\spy).
\end{equation}
Then, employing the condition 2, we can write
\begin{equation}
    S(\be;0)=e^{-iT E_{vac}} \sqrt{2k_{0\be}} \int d\spx\int_{-T/2}^{T/2} dy \vf^*_{\be i}(\spx) G^{(+)}_{ij}(T/2,\spx;y^0,\spy)j^j(y).
\end{equation}
Consequently, in the first Born approximation, the probability to record one photon in the process \eqref{0g} is written as
\begin{equation}\label{probab1}
    dP(\be)=2k_{0\be}\Big| \int d\spx\int_{-T/2}^{T/2} dy \vf^*_{\be i}(\spx)G^{(+)}_{ij}(T/2,\spx;y^0,\spy) j^j(y)\Big|^2 d\be,
\end{equation}
where $d\be$ is the measure in the space of quantum numbers of twisted photons in a vacuum (see formula (14) of \cite{BKL2}). Since
\begin{equation}
    G^{(+)}_{ij}(x,y)=G_{ij}(x,y),\qquad x^0>y^0,
\end{equation}
where $G_{ij}(x,y)$ is the Feynman propagator, the positive-frequency Green function can be replaced by the Feynman propagator in \eqref{probab1}.

Formula \eqref{probab1} can be simplified, if one strengthens the condition 1 and demands that $\vf_\be\approx\vf_{0\be}$ everywhere except from a bounded neighborhood of the region $M$. Then, as long as $\Omega$ is unbounded, the contribution of the bounded neighborhood of the region $M$ can be neglected in the scalar product entering into \eqref{amplitude1}, and $\vf_\be$ can be replaced by $\vf_{0\be}$. As a result,
\begin{equation}\label{probab1_2}
    dP(\be)=2k_{0\be}\Big| \int d\spx\int_{-\infty}^{\infty} dy \vf^*_{0\be i}(\spx)G^{(+)}_{ij}(0,\spx;y^0,\spy)j^j(y)\Big|^2 d\be,
\end{equation}
where we have taken the limit $T\rightarrow\infty$. The concrete choice of the instant of time $x^0=0$ in the argument of the positive-frequency Green function in \eqref{probab1_2} is irrelevant as the different choices of $x^0$ just result in a common phase factor which disappears in evaluating the modulus. Performing the Fourier transforms
\begin{equation}\label{Four_trans}
    G^{(+)}_{ij}(k_0;\spx,\spy)=-2\pi i\sum_\al\de(k_0-k_{0\al}) f_\al\psi_{\al i}(k_0;\spx) \psi^*_{\al j}(k_0;\spy),\qquad j^i(k_0;\spx)=\int_{-\infty}^\infty dt e^{i k_0 t}j^i(t,\spx),
\end{equation}
we obtain
\begin{equation}\label{probab2}
    dP(\be)=2k_{0\be}\Big| \int d\spx d\spy\int_0^\infty\frac{d k_0}{2\pi} \vf^*_{0\be i}(\spx)G^{(+)}_{ij}(k_0;\spx,\spy)j^j(k_0;\spy)\Big|^2 d\be.
\end{equation}
For $k_0>0$, formula \eqref{Four_trans} is reduced to
\begin{equation}
    G^{(+)}_{ij}(k_0;\spx,\spy)=2i\sum_\al \psi_\al(k_0;\spx)\psi^*_\al(k_0;\spy) \im \big[k_0^2-\chi^2_\al(k_0)\big]^{-1}_{k_0\rightarrow k_0+i0}.
\end{equation}
Such a representation of the Fourier transform of the positive-frequency Green function allows one to use formula \eqref{probab2} in the case of a weakly absorbing medium (cf. \cite{GrichSad}). The Green function for the Maxwell equations \eqref{Max_eqns} can be found perturbatively treating $k_0^2(\e(k_0,\spx)-1)$ as a perturbation.

Since the model we consider is exactly solvable (see, e.g., \cite{Glaub2,BaiKatFad,BKL2,GottfYan}), we can find the average number of created twisted photons and the probability of the inclusive process
\begin{equation}\label{inclus_proc}
    0\rightarrow \ga+X.
\end{equation}
The procedure is completely the same as that given in \cite{BKL2}, Sec. 3. The probabilities \eqref{probab1}, \eqref{probab2} coincide with the average number of twisted photons and, with good accuracy, are equal to the probability of inclusive process \eqref{inclus_proc} (see the details in \cite{BKL2}). If the probability to record a twisted photon with quantum numbers $\be\in D$ is needed, then
\begin{equation}
    w_{\text{incl}}(\be\in D;0)=1-\exp\Big[-\int_DdP(\be)\Big].
\end{equation}
For the radiation of a point charged particle, formula \eqref{probab2} looks as
\begin{equation}\label{probab_point}
    dP(\be)=2k_{0\be} e^2\Big| \int_{-\infty}^\infty d\tau \int d\spx \int_0^\infty\frac{d k_0}{2\pi} \vf^*_{0\be i}(\spx)G^{(+)}_{ij}(k_0;\spx,\spx(\tau)) \dot{x}^j(\tau) e^{i k_0 x^0(\tau)}\Big|^2 d\be,
\end{equation}
where $e$ is the particle charge and $x^\mu(\tau)$ specifies the particle worldline. Recall that we work in the system of units where $e^2=4\pi\al$.

\section{Radiation of twisted photons in a homogeneous medium}\label{Homogeneous_Medium}

Let us consider separately the generation of twisted photons in the case when the dispersive medium is homogeneous $\e=\e(k_0)$, $\e(k_0)>0$, and the detector of twisted photons is located in the medium, or the twisted photons escape the medium preserving its phase front structure and are recorded by the detector out of the medium. In that case, the procedure developed above can be considerably simplified.

Denote as $\psi_\al$ the orthonormal set of eigenfunctions of the Maxwell Hamiltonian
\begin{equation}
    \hat{h}\psi_\al=s \tilde{\chi}_\al\psi_\al,\qquad \tilde{\chi}_\al>0,\qquad s=\pm1,
\end{equation}
taken in the form of twisted photons (see the notation in \cite{BKL2}, Sec. 2)
\begin{equation}\label{tw_phot_vac}
\begin{gathered}
    \psi_3(m,k_3,k_\perp)=\frac{1}{\sqrt{RL_z}}j_m(k_\perp x_+,k_\perp x_-) e^{ik_3x_3},\qquad\psi_\pm(s,m,k_3,k_\perp)=\frac{ik_\perp}{sk_0\pm k_3}\psi_3(m\pm1,k_3,k_\perp),\\
    \psi(s,m,k_3,k_\perp)= \frac12[\psi_-(s,m,k_3,k_\perp)\spe_++\psi_+(s,m,k_3,k_\perp)\spe_-]+\psi_3(m,k_3,k_\perp)\spe_3,
\end{gathered}
\end{equation}
where $s$ is the photon helicity, $m$ is the projection of the total angular momentum onto the axis $3$,
\begin{equation}\label{mass_shell_plate1}
    k_0:=\sqrt{k_3^2+k_\perp^2},
\end{equation}
and $R$ and $L_z$ characterize the normalization volume. This set of functions is complete in the space of divergence-free square-integrable complex vector fields. The Maxwell equations \eqref{Max_eqns} entail the on-shell condition
\begin{equation}\label{mass_shell_hom}
    k^2_{0\al}\e(k_{0\al})=\tilde{\chi}^2_\al,\qquad k_{0\al}>0.
\end{equation}
The free quantum electromagnetic field has the form \eqref{em_qfield} with the normalization coefficients
\begin{equation}
    f_\al^{-1}=[k_0^2-\tilde{\chi}^2_\al/\e(k_0)]'_{k_0=k_{0\al}}=[k_0^2\e(k_0)]'_{k_0=k_{0\al}} =\frac{d\tilde{\chi}_\al^2}{dk_{0\al}}.
\end{equation}
In the last equality, it is assumed that $\tilde{\chi}_\al$ is expressed though $k_{0\al}$ by means of the on-shell condition \eqref{mass_shell_hom}. As long as (\cite{LandLifshECM}, Sec. 84)
\begin{equation}
    \frac{d}{dk_0}(k_0^2\e(k_0))>\frac{d}{dk_0}k_0^2>0,
\end{equation}
for $k_0>0$, then $f_\al>0$ for $k_{0\al}>0$.

Repeating the considerations presented in the previous sections (see also \cite{BKL2}), we find that the probability to record a twisted photon produced by point charged particles is given by
\begin{multline}\label{probab_point_uni}
    dP(s,m,k_3,k_\perp)= \frac{k_\perp^3f_\al}{(2\tilde{\chi}_\al)^2}\frac{dk_3dk_\perp}{2\pi^2} \Big|\sum_l e_l\int d\tau_l e^{-ik_{0\al}x^0_l(\tau_l)+ik_3x_{l3}(\tau_l)}\times\\
    \times\Big\{\frac12 \big[ \dot{x}_{l+}(\tau_l)a_-(s,m,k_3,k_\perp;\spx_l(\tau_l)) +\dot{x}_{l-}(\tau_l)a_+(s,m,k_3,k_\perp;\spx_l(\tau_l)) \big]+\dot{x}_{l3}(\tau_l)a_3(m,k_\perp;\spx_l(\tau_l)) \Big\} \Big|^2,
\end{multline}
where the notation borrowed from \cite{BKL2} has been used, $l$ numbers the particles with charges $e_l$, and
\begin{equation}
    \tilde{\chi}_\al=\sqrt{k_3^2+k_\perp^2}.
\end{equation}
In formula \eqref{probab_point_uni}, it is convenient to pass from the variable $k_3$ to $k_0$, $k_0>0$. Then
\begin{equation}
    f_\al dk_3=dk_0/(2k_3),\qquad k_3=\sqrt{k_0^2\e(k_0)-k_\perp^2},
\end{equation}
and
\begin{multline}\label{probab_point_uni1}
    dP(s,m,k_0,k_\perp)= \frac{n^3_\perp}{n_3}\frac{dk_0dk_\perp}{16\pi^2} \Big|\sum_l e_l\int d\tau_l e^{-ik_{0}x^0_l(\tau_l)+ik_3x_{l3}(\tau_l)}\times\\
    \times\Big\{\frac12 \big[ \dot{x}_{l+}(\tau_l)a_-(s,m,k_3,k_\perp;\spx_l(\tau_l)) +\dot{x}_{l-}(\tau_l)a_+(s,m,k_3,k_\perp;\spx_l(\tau_l)) \big]+\dot{x}_{l3}(\tau_l)a_3(m,k_\perp;\spx_l(\tau_l)) \Big\} \Big|^2,
\end{multline}
where
\begin{equation}
    n_3:=\frac{k_3}{k_0\e^{1/2}(k_0)}=\Big(1-\frac{k_\perp^2}{k_0^2\e(k_0)}\Big)^{1/2},\qquad n_\perp:=\frac{k_\perp}{k_0 \e^{1/2}(k_0)}.
\end{equation}
As is seen, the effect of a medium is reduced in this case to the replacement $k_0\rightarrow k_0\e^{1/2}(k_0)$ in the expression for the probability of twisted photon radiation by a classical current in a vacuum \cite{BKL2}.

\paragraph{Vavilov-Cherenkov radiation.}

As an immediate application of general formula \eqref{probab_point_uni1}, we consider the generation of twisted photons by means of the VCh radiation by a charged particle moving strictly along the detector axis. In that case,
\begin{equation}\label{part_traj}
    x_\pm=0,\qquad x_3=\be_3 x^0,
\end{equation}
where $\be_3=const>0$ is the particle velocity in the laboratory frame. The expression under the modulus sign in \eqref{probab_point_uni1} is equal to
\begin{equation}
    2\pi e\be_3\de(k_0-k_3(k_0,k_\perp)\be_3)\de_{m0}.
\end{equation}
Squaring this expression, we obtain the probability of radiation of a twisted photon per unit time
\begin{equation}\label{probab_VCh}
    dP(s,m,k_0,k_\perp)/T=2\pi e^2\de(k_0-k_3(k_0,k_\perp)\be_3)\de_{m0}\be_3^2\frac{n_\perp^3}{n_3}\frac{dk_0dk_\perp}{16\pi^2}.
\end{equation}
We see that the VCh radiation consists of the twisted photons with $m=0$ \cite{IvSerZay,Kaminer16}. This property is a consequence of the general statement \cite{BKL3} that the current density $j^i(t,\spx)$ invariant with respect to rotations around the detector axis for any $t$ produces the twisted photons only with $m=0$. The delta-function entering into \eqref{probab_VCh} defines the Cherenkov cone
\begin{equation}
    k_\perp=k_0\sqrt{\e(k_0)-\be_3^{-2}},\qquad n_\perp=\sqrt{1-\e^{-1}(k_0)\be_3^{-2}},\qquad n_3=\e^{-1/2}(k_0)\be_3^{-1}.
\end{equation}
Recall that the speed of light in a medium is $\e^{-1/2}(k_0)$. Integrating \eqref{probab_VCh} over $k_\perp$, we have
\begin{equation}\label{probab_VCh1}
    dP(s,m,k_0)/T=\frac{\al}{2}\de_{m0}\be_3 \big(1-\e^{-1}(k_0)\be_3^{-2}\big) \theta\big(1-\e^{-1}(k_0)\be_3^{-2}\big) dk_0.
\end{equation}
This expression does not depend on $s$.

\begin{figure}[tp]
\centering
\includegraphics*[align=c,width=0.48\linewidth]{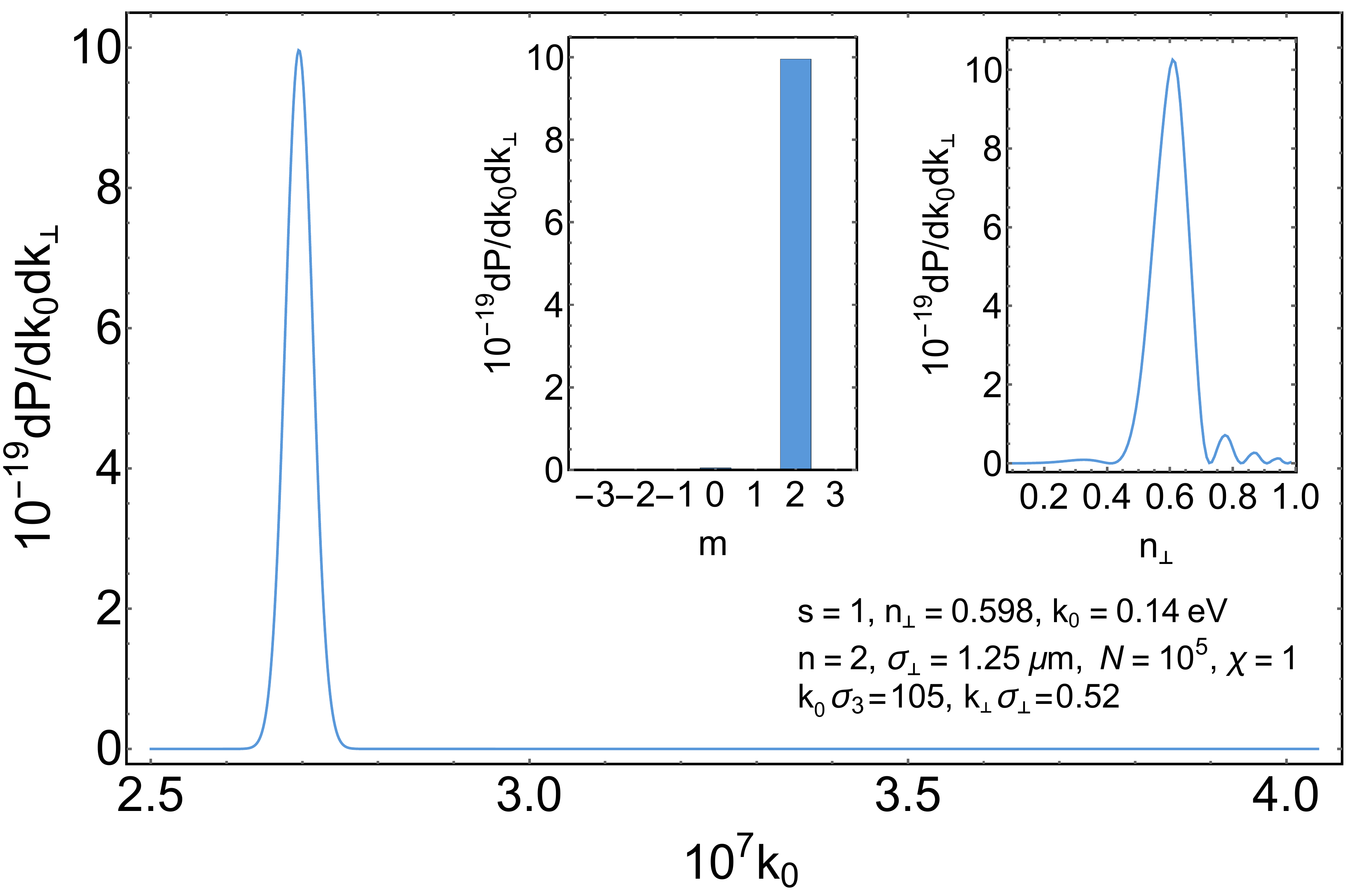}\;
\includegraphics*[align=c,width=0.47\linewidth]{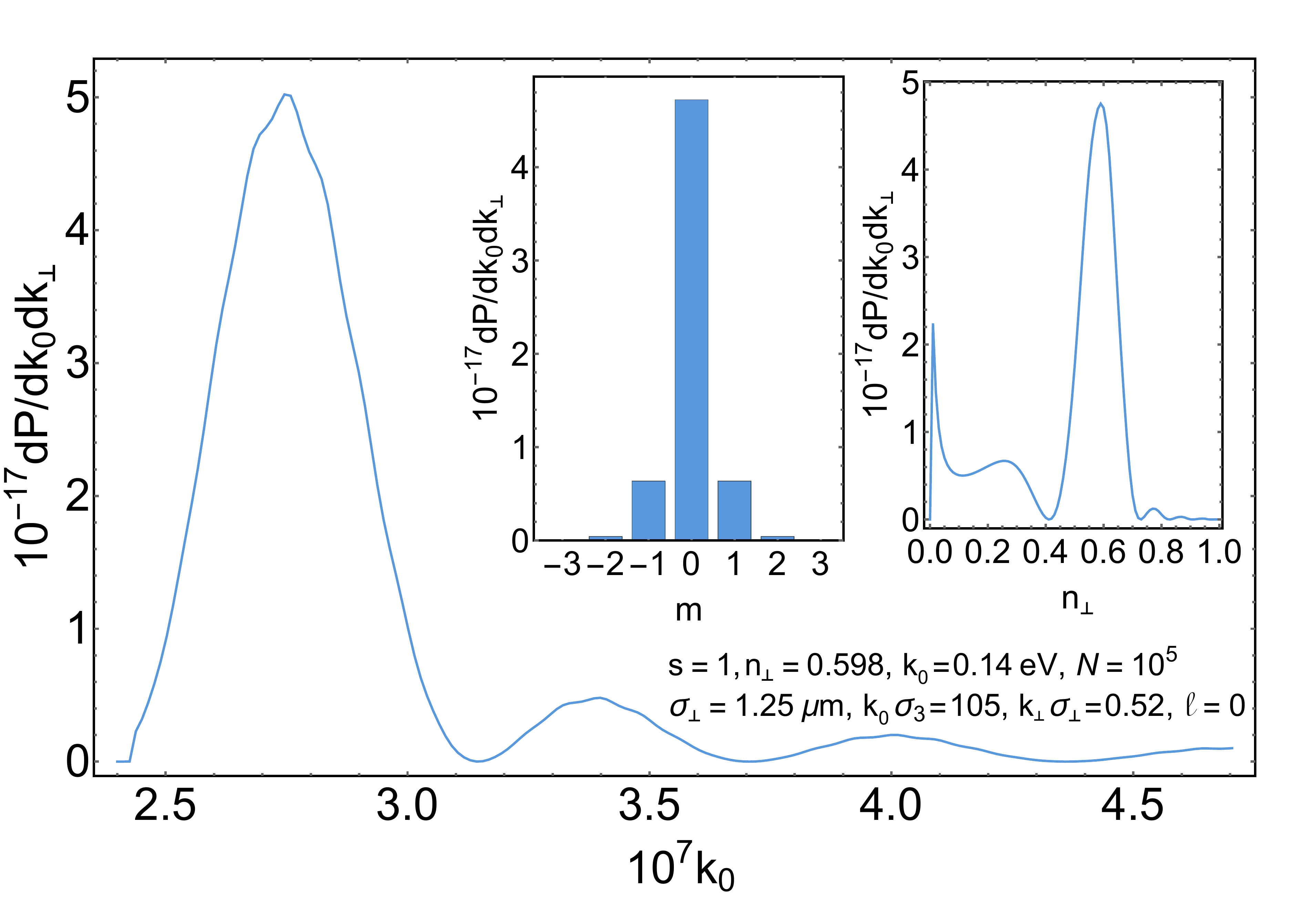}
\caption{{\footnotesize The VCh radiation of twisted photons produced by beams of charged particles in the LiF plate of the thickness $L=100$ $\mu$m. The number of particles in the beam is $N=10^5$ and the Lorentz factor of particles is $\ga=235$. The beam is supposed to have a Gaussian profile (see Appendix \ref{Inter_Fact_App}) with the longitudinal dimension $\s_3=150$ $\mu$m (duration $0.5$ ps) and the transverse size $\s_\perp=1.25$ $\mu$m. The other parameters of the beam are the same as in \cite{BKLb}, Fig. 1. The observation photon energy $k_0=0.14$ eV, which corresponds to the wavelength $9$ $\mu$m. The photon energy depicted on the axis is measured in the electron rest energies, $0.511$ MeV. It is assumed that the particles move along the detector axis towards the detector. The peak at $n_\perp=0.598$ corresponds to the VCh radiation and is found from the standard expression $n_\perp\equiv\sin\theta_{\text{VCh}}=\sqrt{\e'(k_0)-\be^{-2}}$, where $\e'(k_0)$ is the real part of the permittivity. The experimental data for the permittivity $\e(k_0)$ of LiF are taken from \cite{RefrIndex}. The projection of the total angular momentum per photon is denoted as $\ell$. Left panel: The VCh radiation produced by the helically microbunched beam of particles at the second coherent harmonic $k_0=2\pi\chi n\be_3/\de$, where $n=2$ and $\de$ is the helix pitch \cite{BKLb}. The fulfillment of the strong addition rule is evident since the one-particle radiation is concentrated at $m=0$. Right panel: The VCh radiation produced by the uniform Gaussian beam of particles. The coherent contribution to radiation is negligible. As long as $k_\perp\sigma_\perp<1$, the radiation is concentrated near $m=0$ as in the one-particle case \cite{BKb,BKLb}. The peak at the small $n_\perp\sim1/\gamma$ is the transition radiation.}}
\label{diel_plate_fig}
\end{figure}

Using \eqref{probab_VCh1} and the general formulas obtained in \cite{BKb,BKLb}, it is not difficult to find the probability per unit time of twisted photon radiation in the case when the charged particle moves with constant velocity along the line parallel to the detector axis,
\begin{equation}\label{probab_VCh2}
    dP(s,m,k_0)/T=\frac{\al}{2}J^2_m\big(k_\perp(k_0)|x_+|\big) \be_3 \big(1-\e^{-1}(k_0)\be_3^{-2}\big) \theta\big(1-\e^{-1}(k_0)\be_3^{-2}\big) dk_0,
\end{equation}
where $|x_+|$ is the distance from the detector axis to the particle trajectory. Notice that the dependence of expression \eqref{probab_VCh2} on $m$ is the same as for the edge radiation for a charged particle moving along the detector axis \cite{BKL3}. As for the radiation by a bunch of $N$ identical particles moving along parallel trajectories, we obtain
\begin{multline}
    dP_\rho(s,m,k_0)/T=\frac{\al}{2}\big[N f_m\big(k_\perp(k_0)\s_\perp\big) +N(N-1)|\vf_{m}(k_0/\be_3,k_\perp(k_0)\s_\perp)|^2 \big]\times\\
    \times\be_3 \big(1-\e^{-1}(k_0)\be_3^{-2}\big) \theta\big(1-\e^{-1}(k_0)\be_3^{-2}\big) dk_0,
\end{multline}
where $\s_\perp$ is the transverse size of the particle bunch, $f_m(x)$ is the incoherent interference factor, and $\vf_m$ is the corresponding coherent interference factor (see the notation and explicit expressions in \cite{BKb,BKLb}).

\section{Examples}\label{Examples}

\subsection{Dielectric plate}\label{Diel_Plate}

Let us consider the radiation of twisted photons produced by a charged particle passing through a homogeneous isotropic dielectric plate with the width $L$. We suppose that the plate is positioned at $z\in[-L,0]$, and $\e(k_0)>0$ does not depend on the choice of a point in the plate. Out of the plate, $\e(k_0)=1$. The detector of twisted photons is located at $z>0$. The charged particle moves along the detector axis with constant velocity \eqref{part_traj}. The general procedure expounded in Secs. \ref{Quant_in_Medium}, \ref{Recording_Tw_Phot} cannot be directly applied to this case as the medium occupies an infinite volume. Nevertheless, this procedure is readily generalized to such a configuration. Having suitably defined the modes of twisted photons, we shall find the expression for the probability of excitation of these modes by constructing the complete set of solutions of the Maxwell equations \eqref{Max_eqns1} with the boundary conditions \eqref{bound_conds}. Notice that transition radiation of plane-wave photons produced by twisted electrons was investigated in \cite{IvanKarl13l,IvanKarl13a}. In this section, we consider transition radiation of twisted photons generated by usual plane-wave charged particles or by helical beams of them.

The wave functions of twisted photons in a vacuum have the form \eqref{tw_phot_vac}. The wave functions of twisted photons with energy $k_0$ in a homogeneous isotropic medium, $z\in[-L,0]$, are written as
\begin{equation}
    \psi'_3(m',k'_3,k'_\perp)=\psi_3(m',k'_3,k'_\perp),\qquad\psi'_\pm(s',m',k'_3,k'_\perp)=\frac{ik'_\perp}{s'\e^{1/2}(k_0)k_0\pm k'_3}\psi'_3(m'\pm1,k'_3,k'_\perp),
\end{equation}
where
\begin{equation}\label{mass_shell_plate2}
    \e^{1/2}(k_0)k_0=\sqrt{k_3'^2+k_\perp'^2},
\end{equation}
and we have used the notation \eqref{tw_phot_vac}. Notice that
\begin{equation}
    \hat{h}\psi'(s',m',k'_3,k'_\perp)=s'\e^{1/2}(k_0)k_0\psi'(s',m',k'_3,k'_\perp).
\end{equation}
It follows from the first boundary condition in \eqref{bound_conds} at $z=0$ that
\begin{equation}
    k_\perp=k'_\perp,\qquad m=m'.
\end{equation}
For given $k_0$ and $k_3$, the momentum of photon in a medium, $k_3'$, is found from Eq. \eqref{mass_shell_plate2} and can take two values differing by the sign. For $z>0$, the mode function of a twisted photon reads
\begin{equation}\label{vac_mode_f}
    a\psi(s,m,k_3,k_\perp).
\end{equation}
For $z\in[-L,0]$, using the boundary conditions \eqref{bound_conds}, we obtain the wave function
\begin{equation}\label{med_mode_f}
    a\big[b_+\psi'(1,m,k'_3,k_\perp) +c_+\psi'(1,m,-k'_3,k_\perp) +b_-\psi'(-1,m,k'_3,k_\perp) +c_-\psi'(-1,m,-k'_3,k_\perp)\big],
\end{equation}
where the coefficients of the linear combination are
\begin{equation}\label{bpm_cpm}
    b_\pm=\frac{\e^{1/2}\pm s}{4\e k'_3}(\pm sk'_3+\e^{1/2}k_3),\qquad c_\pm=\frac{\e^{1/2}\pm s}{4\e k'_3}(\pm sk'_3-\e^{1/2}k_3).
\end{equation}
The constant $a$ is found from the normalization condition for the mode functions.

\begin{figure}[tp]
\centering
\includegraphics*[align=c,width=0.48\linewidth]{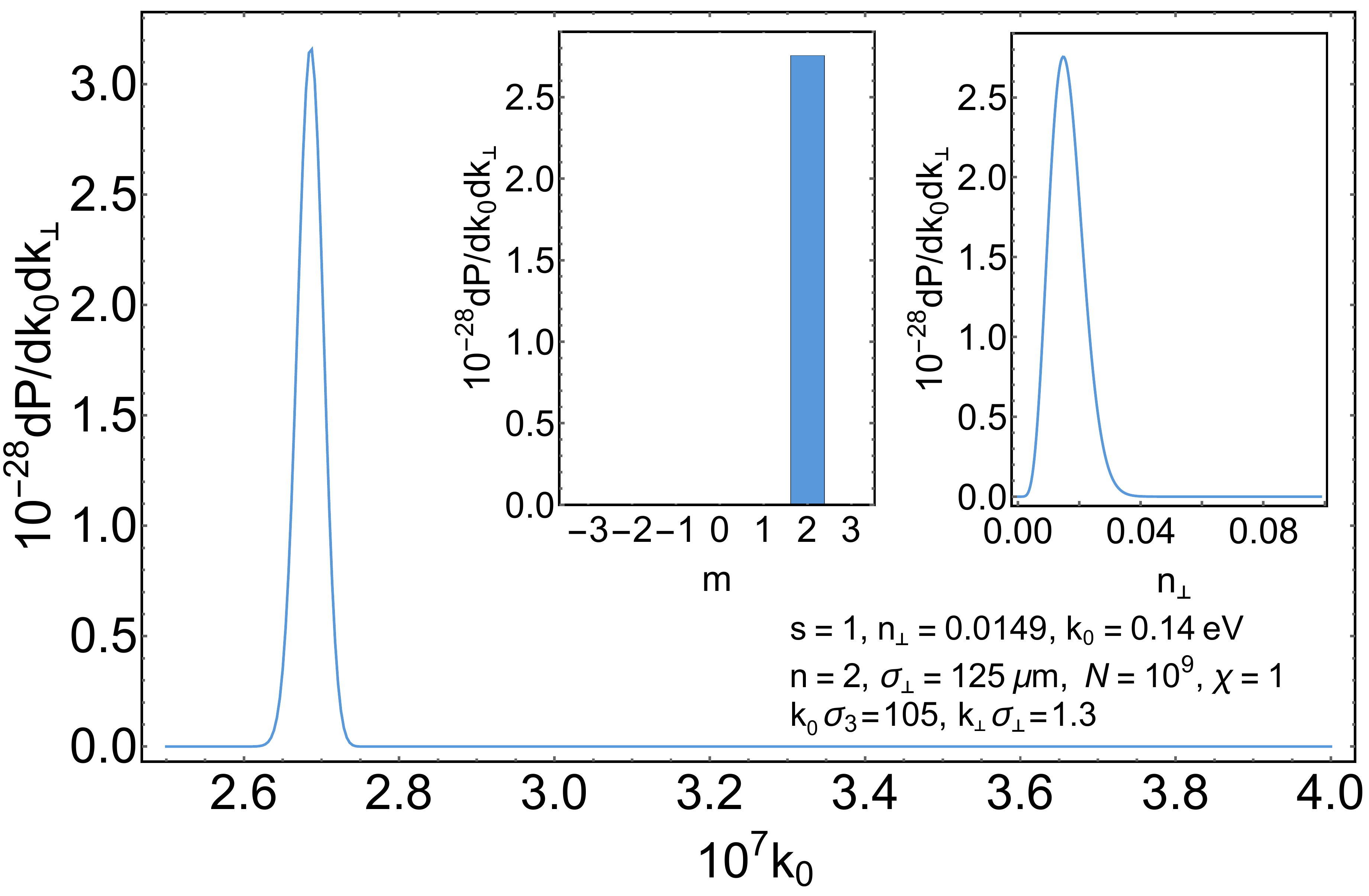}\;
\includegraphics*[align=c,width=0.47\linewidth]{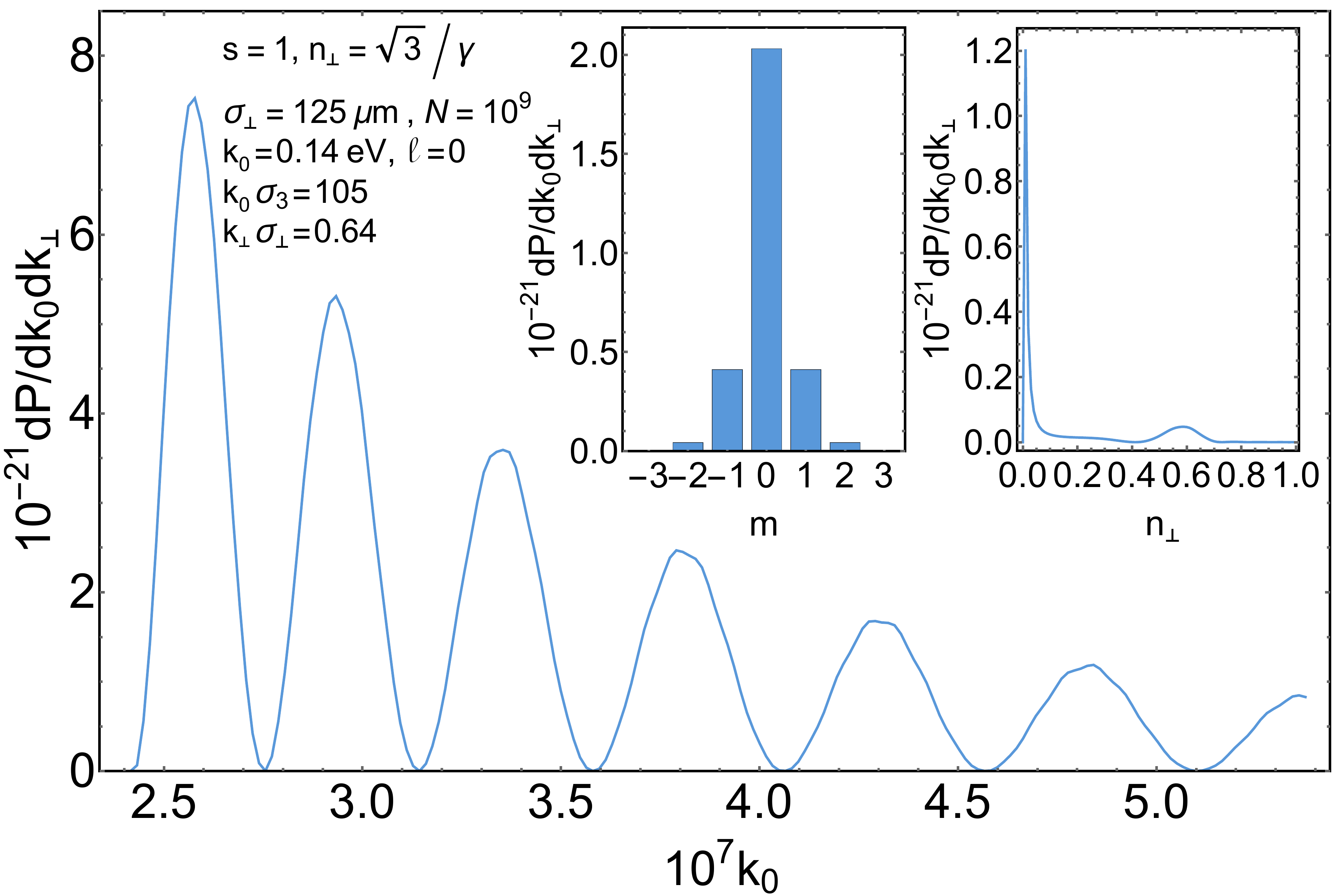}
\caption{{\footnotesize The transition radiation from charged particles traversing the LiF dielectric plate. The particles move along the detector axis towards the detector. The parameters of the plate, the profile of the beam, and the Lorentz factor of particles are the same as in Fig. \ref{diel_plate_fig}. Left panel: The transition radiation of the helically microbunched beam of particles. The fulfillment of the strong addition rule on the second coherent harmonic of radiation is clearly seen. Right panel: The transition radiation of the uniform Gaussian beam of particles. The contribution of coherent radiation is strongly suppressed. The small hump at $n_\perp\approx0.6$ is the VCh radiation.}}
\label{diel_plate2_fig}
\end{figure}

For $z<-L$, the mode function is the linear combination
\begin{equation}\label{vac_mode_f2}
    a\big[a_+\psi(1,m,k_3,k_\perp) +d_+\psi(1,m,-k_3,k_\perp) +a_-\psi(-1,m,k_3,k_\perp) +d_-\psi(-1,m,-k_3,k_\perp)\big].
\end{equation}
Then the boundary conditions at $z=-L$ lead to
\begin{equation}\label{Fresnel_coeff}
\begin{split}
    a_{\pm}&=\frac{2 (1\pm s)\e k_3k'_3\cos(k'_3L)-i(\e^{2}k_{3}^{2}+k'^2_{3}\pm s\e(k_{3}^{2}+k'^2_{3}))\sin(k'_3L)}{4\e k_3k'_3}e^{ik_3L}, \\
    d_{\pm}&=-i\frac{\e^{2}k_{3}^{2}-k'^2_{3}\pm s\e(k_{3}^{2}-k'^2_{3})}{4\e k_3k'_3}\sin(k'_3L) e^{-ik_3L},
\end{split}
\end{equation}
where $s$ is the helicity of the mode function \eqref{vac_mode_f}. The coefficients \eqref{Fresnel_coeff} obey the unitarity relation
\begin{equation}
    1+|d_+|^2+|d_-|^2=|a_+|^2+|a_-|^2,
\end{equation}
for real $k_3$. The dielectric plate is ideally transparent for the modes with real $k_3$ and $k'_3L=\pi n$, $n\in \mathbb{Z}$.

Let $\Phi(s,m,k_\perp,k_3)$ be the mode function taking the values \eqref{vac_mode_f}, \eqref{med_mode_f}, and \eqref{vac_mode_f2} on the corresponding intervals of the variable $z$ without the common factor $a$ and with the same values of the quantum numbers $s$, $m$, $k_\perp$, and $k_0$. The various linear combinations of the modes $\Phi(s,m,k_\perp,k_3)$ with the same energy give the complete set of solutions to the Maxwell equations \eqref{Max_eqns1} with the boundary conditions \eqref{bound_conds}. There are bound states among these mode functions. They are exponentially damped out of the dielectric plate and correspond to the case of complete internal reflection of the electromagnetic wave in the dielectric. This occurs when
\begin{equation}\label{intern_relec}
    k'^2_3=\e(k_0) k_0^2-k_\perp^2\geqslant0,\qquad k_3^2=k_0^2-k_\perp^2<0.
\end{equation}
and is only possible for $\e>1$. Setting $k_3=i|k_3|$, we find that the linear combination,
\begin{equation}\label{bound_state}
    \al_+\Phi(1,m,k_\perp,k_3)+\al_-\Phi(-1,m,k_\perp,k_3),
\end{equation}
is exponentially damped out of the dielectric plate if one of the following two conditions is satisfied:
\begin{equation}
    \ctg (k'_3L)=\frac12\Big(\frac{k'_3}{|k_3|}-\frac{|k_3|}{k'_3}\Big),\qquad \ctg (k'_3L)=\frac12\Big(\frac{k'_3}{\e|k_3|}-\frac{\e|k_3|}{k'_3}\Big).
\end{equation}
These equations define the multivalued function $k_\perp=k_\perp(k_0,L)$. Such states are irrelevant for the further investigation of radiation of twisted photons. They are exponentially suppressed in the domain where the detector is positioned and give a negligible contribution to \eqref{probab1_2}. Notice that the bound states \eqref{bound_state} are the eigenfunctions of the operator of the total angular momentum.

It is useful to numerate the mode functions of twisted photons recorded by the detector ($k_3>0$) by the quantum numbers $\al:=(s,m,k_3,k_\perp)$ and put
\begin{equation}\label{k3p}
    k'_3=\sqrt{\e(k_0) k_0^2-k_\perp^2},\qquad k_0=\sqrt{k_\perp^2+k_3^2}.
\end{equation}
Then
\begin{equation}\label{norm_coeff}
    f^{-1}_\al=2k_{0\al}=2\sqrt{k_3^2+k_\perp^2}>0.
\end{equation}
Since, in the region $z>0$, where the detector is positioned, the wave functions $\Phi(s,m,k_\perp,k_3)$ coincide with the wave functions of twisted photons in a vacuum, we will call these wave functions as the mode functions of twisted photons. For $L_z\rightarrow\infty$ and $R\rightarrow\infty$, the normalization condition for the mode function $a\Phi(s,m,k_\perp,k_3)$ is written as
\begin{equation}
    \frac{|a|^2}{2}(1+|a_+|^2+|a_-|^2+|d_+|^2+|d_-|^2)=|a|^2(|a_+|^2+|a_-|^2)=1.
\end{equation}
The factor $1/2$ in this expression is owning to the fact that the dielectric plate divides all the space into two parts. The wave function of a photon in the plate gives a negligible contribution to the normalization condition when $L_z\rightarrow\infty$, $R\rightarrow\infty$. Substituting the explicit expressions \eqref{Fresnel_coeff}, we deduce
\begin{equation}
    |a|^{-2}=|a_+|^2+|a_-|^2=\Big|1+\frac18\Big[(\e^2+1)\Big(\frac{k^2_3}{k'^2_3} +\frac{k'^2_3}{\e^2 k^2_3}\Big)-4\Big]\sin^2(k'_3 L)\Big|.
\end{equation}
As a result, in accordance with the general formulas obtained in the previous sections, the probability to record a twisted photon produced by the particles with charges $e_l$ looks as
\begin{multline}\label{prob_plate}
    dP(s,m,k_\perp,k_3)=|a|^2\bigg|\sum_{l}e_l\int_{-\infty}^\infty d\tau e^{-ik_0 x^0_l(\tau_l)}
    \Big\{\dot{x}_{3l}(\tau_l)\Phi_3(s,m,k_\perp,k_3;\spx_l(\tau_l))+\\
    +\frac12\big[\dot{x}_{+l}(\tau_l)\Phi_-(s,m,k_\perp,k_3;\spx_l(\tau_l)) +\dot{x}_{-l}(\tau_l)\Phi_+(s,m,k_\perp,k_3;\spx_l(\tau_l))\big] \Big\} \bigg|^2 \Big(\frac{k_\perp}{2k_0}\Big)^3\frac{dk_3 dk_\perp}{2\pi^2},
\end{multline}
where $x_l^\mu(\tau)$ are the particle worldlines.

\begin{figure}[tp]
\centering
\includegraphics*[align=c,width=0.48\linewidth]{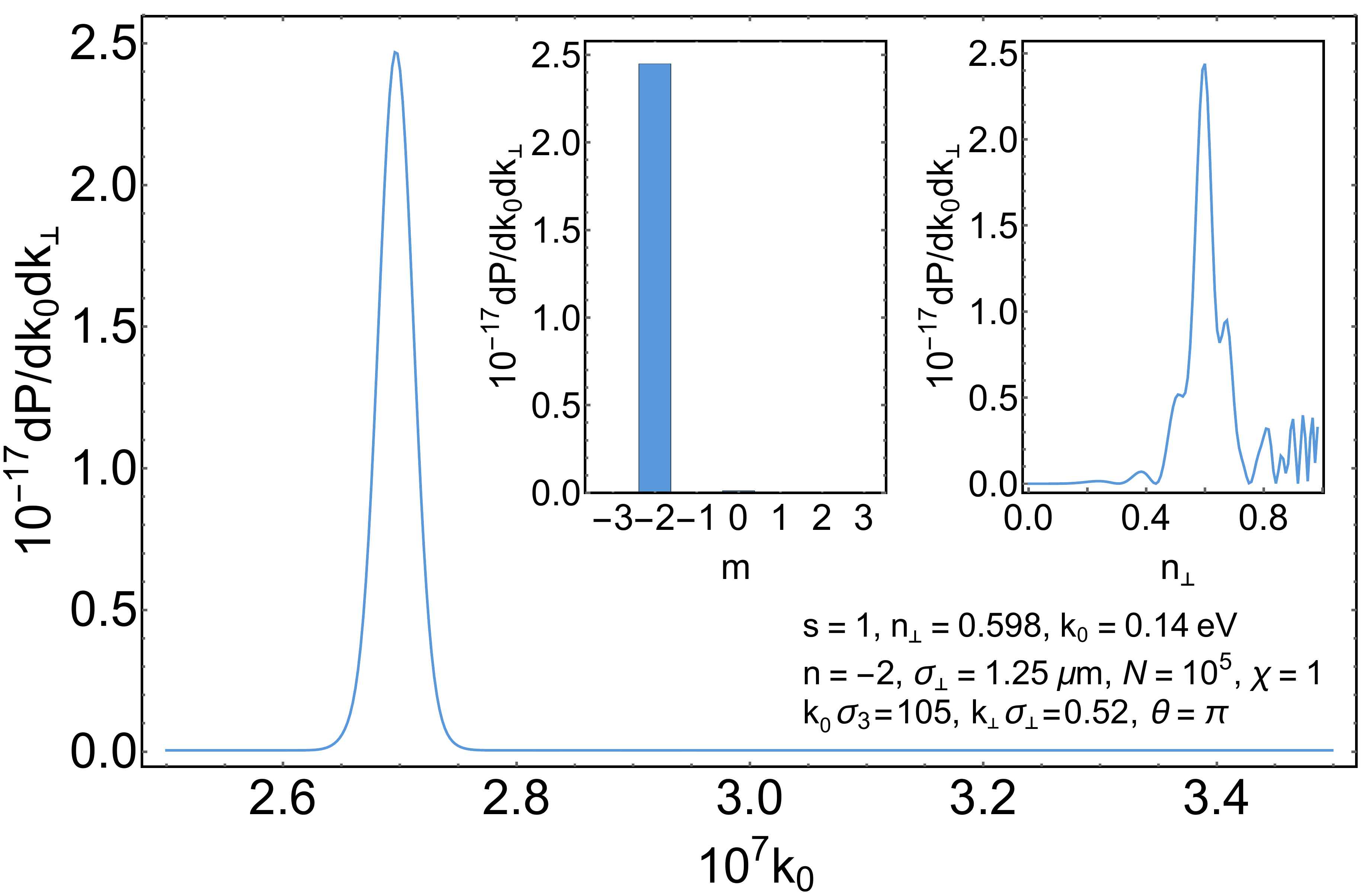}\;
\includegraphics*[align=c,width=0.48\linewidth]{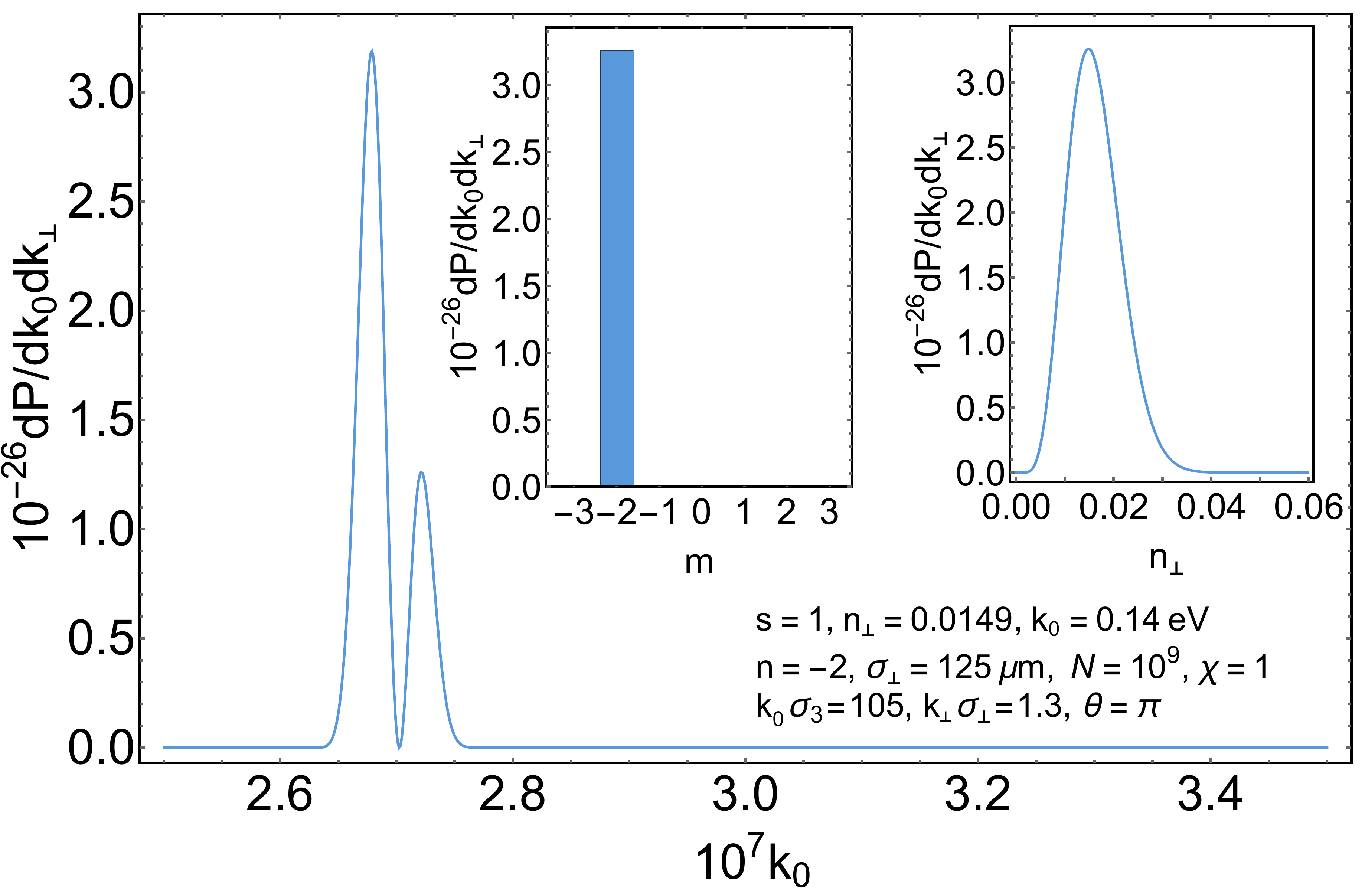}
\caption{{\footnotesize The same as in Figs. \ref{diel_plate_fig}, \ref{diel_plate2_fig}, but the charged particles move from the detector, i.e., $\theta=\pi$. Left panel: The VCh radiation of a helically microbunched beam of particles. In accordance with the strong addition rule, since the beam of particles moves from the detector, the distribution over $m$ is shifted to the opposite direction in comparison with Fig. \ref{diel_plate_fig}. Right panel: The transition radiation of a helically microbunched beam of particles.}}
\label{diel_plate3_fig}
\end{figure}

Considering the particles with the trajectories \eqref{part_traj}, the expression under the modulus sign in \eqref{prob_plate} at $m=0$ is given by
\begin{multline}\label{ampl_plane}
    A:=\frac{i|\be_3|}{k_0}\bigg[\frac{\Big(\cos(k'_3L)-\frac{i}{2}\Big(\frac{\e k_3}{k'_3}+\frac{k'_3}{\e k_3}\Big)\sin(k'_3L)\Big)e^{ik_0L/\be_3}-1}{1-n_3\be_3} +\frac{-\frac{i}{2}\Big(\frac{\e k_3}{k'_3}-\frac{k'_3}{\e k_3}\Big)\sin(k'_3L)e^{ik_0L/\be_3}}{1+n_3\be_3} +\\
    +\frac{1+\frac{\e k_3}{k'_3}}{2\e(1-n'_3\be_3)}\big(1-e^{i k_0L(1-n'_3\be_3)/\be_3}\big) +\frac{1-\frac{\e k_3}{k'_3}}{2\e(1+n'_3\be_3)}\big(1-e^{i k_0L(1+n'_3\be_3)/\be_3}\big) \bigg],
\end{multline}
where $n'_3:=k'_3/k_0$. The probability to record a twisted photon produced by the particle with charge $e$ takes the form
\begin{equation}\label{probab_plate}
    dP(s,m,k_\perp,k_3)=e^2 |a|^2|A|^2\de_{0,m}\Big(\frac{k_\perp}{2k_0}\Big)^3\frac{dk_3 dk_\perp}{2\pi^2},
\end{equation}
This probability is concentrated at $m=0$ and does not depend on the photon helicity. The first two terms in \eqref{ampl_plane} describe transition radiation, and the terms on the second line \eqref{ampl_plane} correspond to VCh radiation \cite{Pafomov,GinzTsyt}. For
\begin{equation}\label{Cherenk_cond}
    k_0L|1\mp n'_3\be_3|/|\be_3|\lesssim\pi/5,
\end{equation}
the latter contributions possess a sharp maximum of the order
\begin{equation}
    e^2L^2 |a|^2\frac{|1\pm \frac{\e k_3}{k'_3}|^2}{4|\e|^2}\de_{0,m} \Big(\frac{k_\perp}{2k_0}\Big)^3\frac{dk_3 dk_\perp}{2\pi^2},
\end{equation}
where the sign ``$\pm$'' agrees with the choice of the sign in \eqref{Cherenk_cond}. The contribution of transition radiation to \eqref{probab_plate} reaches a maximum at
\begin{equation}\label{n_perp_max}
    n_\perp\ga\approx\sqrt{3},
\end{equation}
provided $k_0L/\ga^2\ll1$. For $\be_3>0$, formula \eqref{probab_plate} describes the radiation of a charged particle moving towards the detector, while for $\be_3<0$ the particle moves from the detector.

\begin{figure}[tp]
\centering
\includegraphics*[align=c,width=0.48\linewidth]{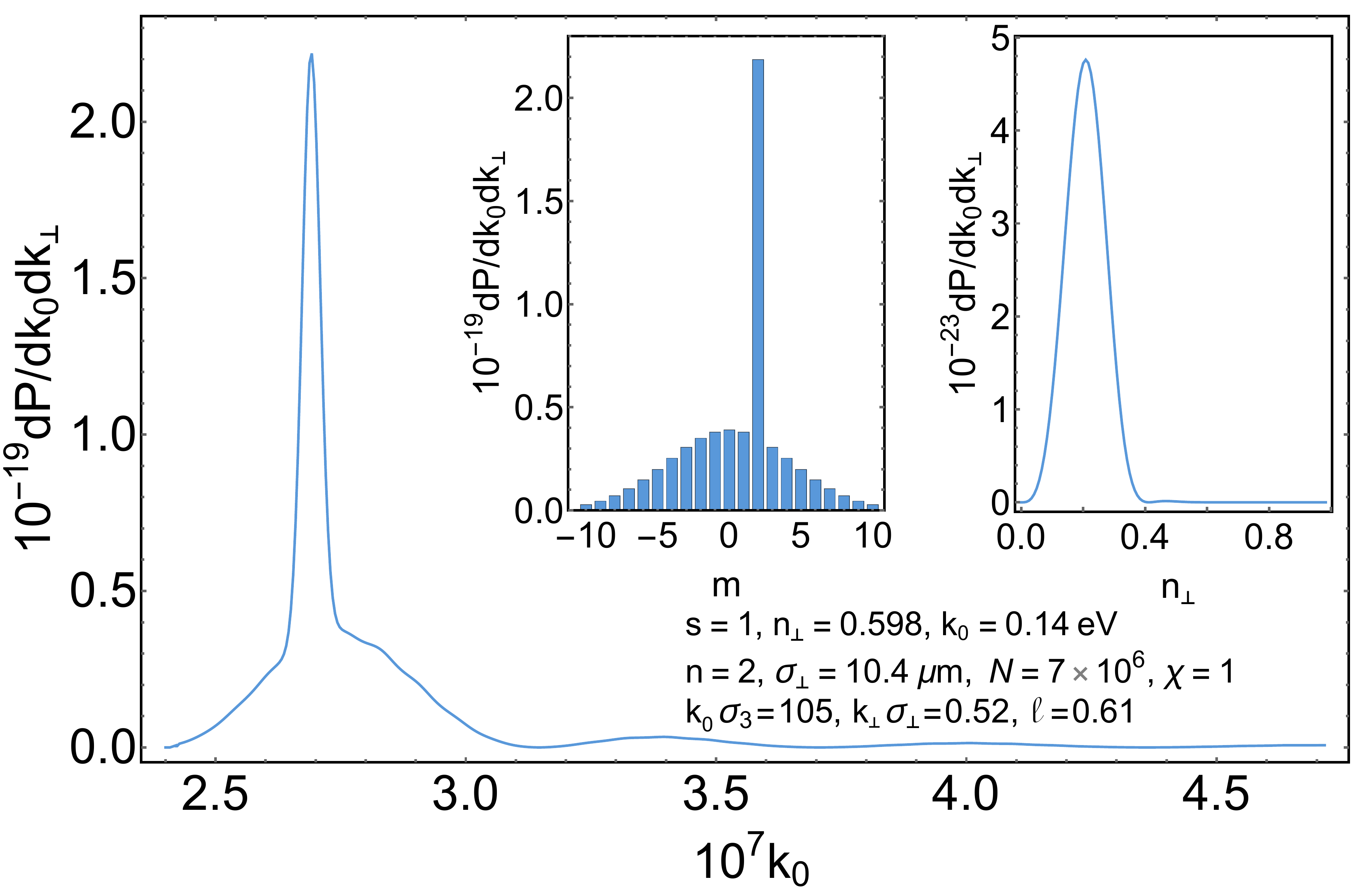}\;
\includegraphics*[align=c,width=0.465\linewidth]{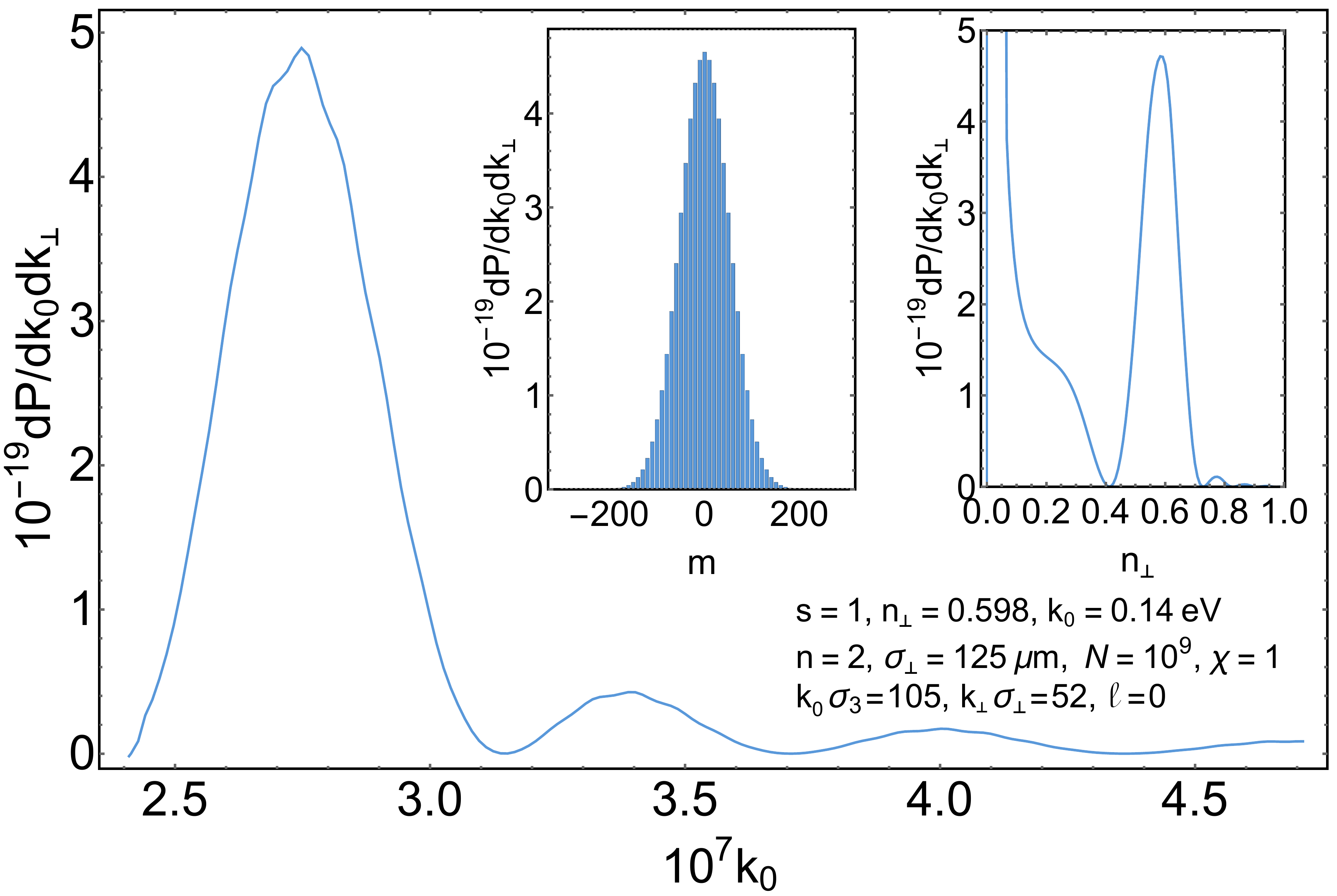}
\caption{{\footnotesize The suppression of the coherent contribution to VCh radiation in increasing the transverse size of the helically microbunched particle beam (cf. Fig. \ref{diel_plate_fig}). The numbers of particles in the beams are such that the densities of particles are equal. Other parameters are the same as in Fig. \ref{diel_plate_fig}. The large peaks at small $n_\perp$ correspond to transition radiation. Left panel: The plots against $k_0$ and $n_\perp$ are given for $m=2$. Right panel: The plots against $k_0$ and $n_\perp$ are given for $m=0$.}}
\label{diel_plate4_fig}
\end{figure}

\begin{figure}[tp]
\centering
\includegraphics*[align=c,width=0.48\linewidth]{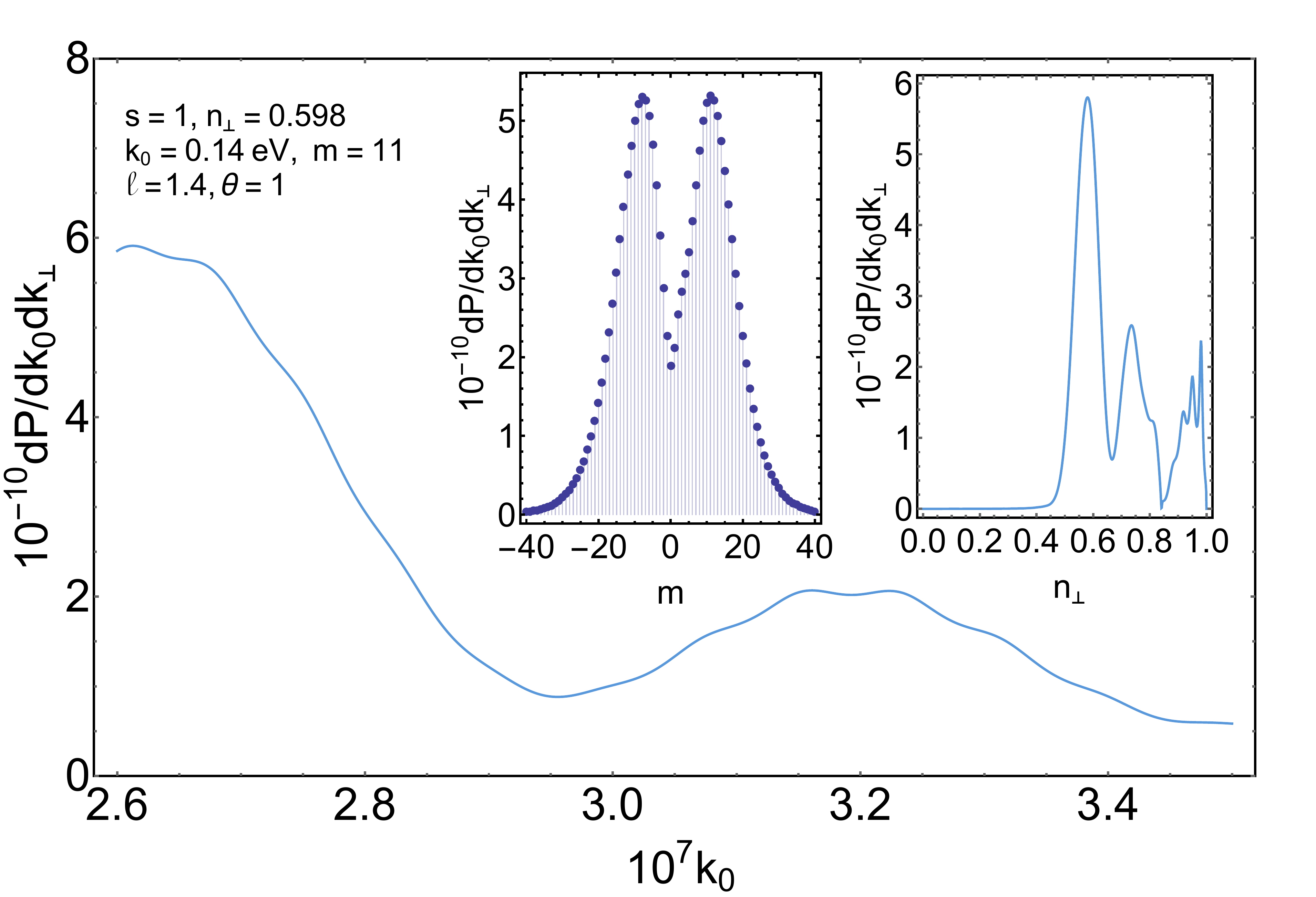}\;
\includegraphics*[align=c,width=0.48\linewidth]{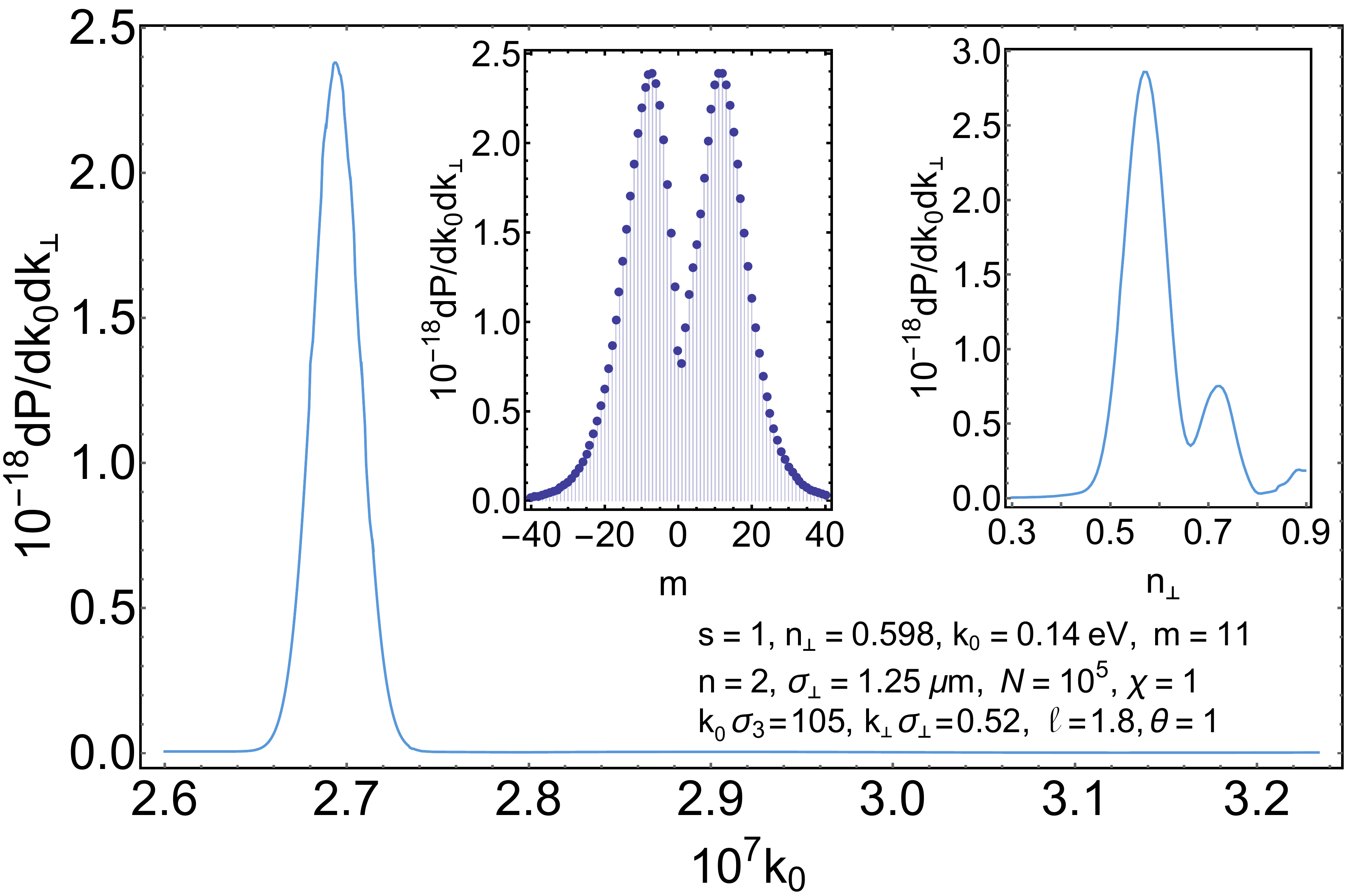}
\caption{{\footnotesize The radiation of twisted photons by charged particles traversing the LiF dielectric plate at the angle $\theta=1$. The angle is counted from the normal to the plate. The other parameters are the same as in Fig. \ref{diel_plate_fig}. Left panel: The radiation of twisted photons produced by one charged particle. The plots against $k_0$ and $n_\perp$ are given at $m=11$. Right panel: The radiation of twisted photons produced by helically microbunched particle beam at the second coherent harmonic \eqref{coher_harmcs}. The plots against $k_0$ and $n_\perp$ are given at $m=11$. Of course, the strong addition rule does not hold, but the harmonic \eqref{coher_harmcs} is clearly seen.}}
\label{diel_plate5_fig}
\end{figure}

The fact that probability \eqref{probab_plate} is concentrated at $m=0$ is a consequence of the general rule \cite{BKL3}. This property holds for any current density symmetric under the rotations around the detector axis. If, instead of one charged particle, the radiation of $N$ identical charged particles moving along parallel trajectories is considered, then, as was shown in \cite{BKb,BKLb}, the probability to record a twisted photon becomes
\begin{equation}\label{prob_rad_tot_hw}
    dP_\rho(s,m,k_\perp,k_3)=\big[Nf_{m}+N(N-1)|\vf_{m}|^2\big]dP_1(s,0,k_\perp,k_3),
\end{equation}
where $dP_1(s,0,k_\perp,k_3)$ is the probability of radiation of a twisted photon by one charged particle, i.e., in our case \eqref{probab_plate}. Of course, it is assumed in \eqref{prob_rad_tot_hw} that the particle beam falls normally onto the dielectric plate. The functions $f_{m}$ and $\vf_m$ are the incoherent and coherent interference factors, respectively. The explicit expressions for $f_{m}$ and $\vf_m$ are given in \cite{BKb,BKLb}. In particular, the probability distribution over $m$ for the coherent radiation of twisted photons by a helically microbunched beam of charged particles moving along parallel trajectories is shifted with respect to the one-particle probability distribution by the signed number $n$ of coherent harmonic (the strong addition rule). This allows one to generate the twisted photons with large $m$ by means of coherent transition and VCh radiations (see Figs. \ref{diel_plate_fig}, \ref{diel_plate2_fig}, \ref{diel_plate3_fig}, \ref{diel_plate4_fig}). In particular, the projection of the total angular momentum per photon, $\ell_\rho$, of the radiation at the $n$-th coherent harmonic is given by \cite{BKLb}
\begin{equation}\label{ang_mom_per_phot}
    \ell_\rho=\ell_1+n\frac{(N-1)|\vf_n|^2}{1+(N-1)|\vf_n|^2},
\end{equation}
where $\ell_1$ is the projection of the total angular momentum per photon for the radiation produced by one charged particle. In our case, $\ell_1=0$. Notice that, in the paraxial regime, $n_\perp\ll1$, even the twisted photons with $m=0$ and definite helicity $s$ possess a nontrivial phase front corresponding to the orbital angular momentum  $l=m-s=-s$ \cite{IvSerZay}. The probability to record a twisted photon radiated by one particle moving along a straight line parallel to the detector axis at a distance $|x_+|$ is obtained from \eqref{prob_rad_tot_hw} by the substitution
\begin{equation}\label{1_pert_prl}
    f_m=J_m^2(k_\perp|x_+|),\qquad N=1.
\end{equation}
The radiation of twisted photons by the helically microbunched particle beam traversing the dielectric plate at a large angle to its normal is described in Fig. \ref{diel_plate5_fig}.

\subsection{Thick dielectric plate}\label{Thick_Diel_Plate}

Now we consider the case when thickness of the dielectric plate is so large that one may suppose that the half-space $z<0$ is filled by the dielectric, i.e., $L\rightarrow\infty$. The mode functions $\Phi(s,m,k_\perp,k_3)$ are given by formulas \eqref{vac_mode_f}, \eqref{med_mode_f} without the common factor $a$. The states exponentially damped in the region $z>0$ correspond to $\Phi(s,m,k_\perp,k_3)$ with the quantum numbers $k_0$, $k_\perp$ satisfying \eqref{intern_relec}. These states are not recorded by the detector located sufficiently far from the medium in the domain $z>0$.

We will be interested in the states with $k_3>0$ undamped in the region $z>0$. It is convenient to number them by  $\al:=(s,m,k_3,k_\perp)$. Then the normalization coefficient takes the form \eqref{norm_coeff}. For $R\rightarrow\infty$, $L_z\rightarrow\infty$, the normalization of the wave functions leads to the relation
\begin{equation}\label{norm_half}
    \frac{|a|^2}{2}\big[1+\e(|b_+|^2+|b_-|^2+|c_+|^2+|c_-|^2)\big]=1,
\end{equation}
where the factor $1/2$ comes from the fact that the whole space is split into two parts by the interface $\Si$ and $\e$ in front of the parenthesis results from the integration measure in the scalar product \eqref{scal_prod}. Substituting the explicit expressions \eqref{bpm_cpm}, we find the normalization factor
\begin{equation}\label{norm_fact}
    |a|^{-2}=\frac12\Big|1+(\e+1)\frac{\e k_3^2+k'^2_3}{4\e k'^2_3}\Big|.
\end{equation}
Inasmuch as the mode functions $\Phi(s,m,k_\perp,k_3)$ coincide with the vacuum twisted photons for $z>0$, we will call them as twisted photons. The probability of radiation of twisted photons by charged particles is described by formula \eqref{prob_plate}.

As for one charged particle moving along the trajectory \eqref{part_traj}, the probability to record a twisted photon produced by it becomes
\begin{equation}
    dP(s,m,k_\perp,k_3)=e^2 |a|^2\frac{\be^2_3}{k_0^2}\bigg|\frac{\frac{1}{2\e}\Big(1+\frac{\e k_3}{k'_3}\Big)}{1-n'_3\be_3} +\frac{\frac{1}{2\e}\Big(1-\frac{\e k_3}{k'_3}\Big)}{1+n'_3\be_3} -\frac{1}{1-n_3\be_3}\bigg|^2\de_{0,m}\Big(\frac{k_\perp}{2k_0}\Big)^3\frac{dk_3 dk_\perp}{2\pi^2},
\end{equation}
where $e$ is the particle charge. The singularity of this expression when the Cherenkov condition is fulfilled is resolved by taking into account the medium absorption, i.e., one has to suppose that $\e(k_0)\in \mathbb{C}$ and $k'_3$ is found from \eqref{k3p}. The last term under the modulus sign describes the contribution of transition radiation. It reaches a maximum when \eqref{n_perp_max}. Due to the symmetry of the problem, the radiation probability is concentrated at $m=0$. Moreover, it does not depend on the photon helicity.

The probability to radiate a twisted photon by a bunch of particles moving along parallel trajectories is given by \eqref{prob_rad_tot_hw}. The radiation of twisted photons by helically microbunched beams of particles obeys the strong addition rule and the sum rule \eqref{ang_mom_per_phot} is satisfied. As for radiation by one particle moving along a straight line parallel to the detector axis, one just needs to make substitution \eqref{1_pert_prl} in \eqref{prob_rad_tot_hw}.

It is not difficult to find the probability to record a twisted photon produced by a charged particle moving along the trajectory
\begin{equation}\label{traj_tilt}
    x_\pm=\be_\pm x^0,\qquad x_3=\be_3 x^0,
\end{equation}
where $\be_\pm$ and $\be_3$ are the constant projections of the particle velocity in the laboratory frame. Let us introduce the notation
\begin{equation}\label{kq}
\begin{split}
    \kappa(n_{3},\be_{3})&:=\sqrt{(1-n_{3}\be_{3})^{2}-n_{\perp}^{2}\be_{\perp}^{2}},\\
    q(n_{3},\be_{3})&:=\frac{n_{\perp}\be_{\perp}}{1-n_{3}\be_{3}+\kappa(n_{3},\be_{3})} =\frac{1-n_{3}\be_{3}-\kappa(n_{3},\be_{3})}{n_{\perp}\be_{\perp}},
\end{split}
\end{equation}
and $\de:=\arg\be_+$. Then the expression under the modulus sign in \eqref{prob_plate} reads as
\begin{equation}\label{A_trans}
    A=I(\be_3)+I'(n'_3)+I'(-n'_3),
\end{equation}
where $I(\be_3)$ is the amplitude of the edge radiation \cite{BKL3}
\begin{equation}\label{I_edge}
\begin{alignedat}{3}
    I(\be_{3})&:=\sgn(\be_3)\frac{i^{-1-m}}{k_{\perp}n_{\perp}}\Big(\frac{\be_{3}-n_{3}}{\kappa(n_{3},\be_3)}-s \sgn(m)\Big) e^{i m \de} q^{|m|}(n_{3},\be_{3}),& \quad \textrm{for} \quad|m|&>0;\\
    I(\be_{3})&:=\sgn(\be_3)\frac{i^{-1}}{k_{\perp}n_{\perp}}\Big(\frac{\be_{3}-n_{3}}{\kappa(n_{3},\be_3)}+ n_{3} \Big),&\quad \textrm{for} \quad|m|&=0;
\end{alignedat}
\end{equation}
while the other two terms give the amplitude of the transition radiation
\begin{equation}\label{I_trans}
\begin{alignedat}{3}
    I'(n'_{3})&:=-\sgn(\be_3)\frac{i^{-1-m}}{2 k_\perp n_\perp} \Big[\frac{\e\be_{3}-n'_{3}}{\kappa(n'_{3},\be_{3})} \Big(\frac{n_3}{n'_3}+\frac{1}{\e}  \Big) -s \sgn(m) \Big(\frac{n_3}{n'_3}+1 \Big) \Big] e^{i m \de} q^{|m|}(n'_{3},\be_{3}),& \quad \textrm{for} \quad|m|&>0;\\
    I'(n'_{3})&:=-\sgn(\be_3)\frac{i^{-1}}{2k_{\perp}n_{\perp}}\Big(\frac{n_3}{n'_3}+\frac{1}{\e}  \Big) \Big[\frac{\e\be_{3}-n'_{3}}{\kappa(n'_{3},\be_{3})}+ n'_{3} \Big],&\quad \textrm{for} \quad|m|&=0.
\end{alignedat}
\end{equation}
The probability of twisted photon radiation is written as
\begin{equation}\label{probab_plate_thck}
    dP(s,m,k_\perp,k_3)=e^2 |a|^2|A|^2\Big(\frac{k_\perp}{2k_0}\Big)^3\frac{dk_3 dk_\perp}{2\pi^2},
\end{equation}
where $|a|^2$ is presented in \eqref{norm_fact}.

It was shown in \cite{BKL3} that the probability of radiation of twisted photons for an arbitrary QED process in a vacuum possesses the reflection symmetry,
\begin{equation}\label{refl_symm}
    dP(s,m,k_\perp,k_3)=dP(-s,-m,k_\perp,k_3),
\end{equation}
in the infrared regime. It is clear from \eqref{A_trans}, \eqref{I_edge}, \eqref{I_trans}, and \eqref{probab_plate_thck} that, in the case we consider, this symmetry also holds. Moreover, this symmetry takes place for the processes involving an arbitrary number of charged particles and evolving near the boundary of the thick dielectric plate. Namely, the probability to record a twisted photon produced by an arbitrary number of charged particles moving along the trajectories \eqref{traj_tilt} with different velocities $\boldsymbol{\be}$ possesses the reflection symmetry \eqref{refl_symm} provided the medium is transparent, i.e., $\e>1$, and
\begin{equation}
    (1-n'_3|\be_3|)^2\geqslant n_\perp^2\be_\perp^2,\qquad n'_3=\sqrt{\e-n_\perp^2},
\end{equation}
for all the particles participating in the process. This property also holds when the particles transmute one into another, are created, or cease to exist at the origin. The proof of the symmetry property \eqref{refl_symm} is completely analogous to that given in \cite{BKL3}.

\subsection{Conducting plate}\label{Cond_Plate}

In order to describe the radiation of twisted photons by charged particles moving near an ideally conducting plane $z=0$, it is necessary to construct the corresponding mode functions in the domain $z>0$ satisfying the boundary condition \eqref{bound_conds_cond}. The linear combination
\begin{equation}
    \Phi(s,m,k_\perp,k_3)=\psi(s,m,k_\perp,k_3)+b_+\psi(1,m,k_\perp,-k_3)+b_-\psi(-1,m,k_\perp,-k_3),\qquad k_3>0,
\end{equation}
satisfies \eqref{bound_conds_cond} when
\begin{equation}
    b_\pm=\frac{1\mp s}{2}.
\end{equation}
The normalization of the wave function $a\Phi(s,m,k_\perp,k_3)$ results in
\begin{equation}
    \frac{|a|^2}{2}(1+|b_+|^2+|b_-|^2)=|a|^2=1.
\end{equation}
We will call these mode functions as twisted photons since the wave traveling to the detector coincides with the vacuum twisted photon.

\begin{figure}[tp]
\centering
\includegraphics*[align=c,width=0.48\linewidth]{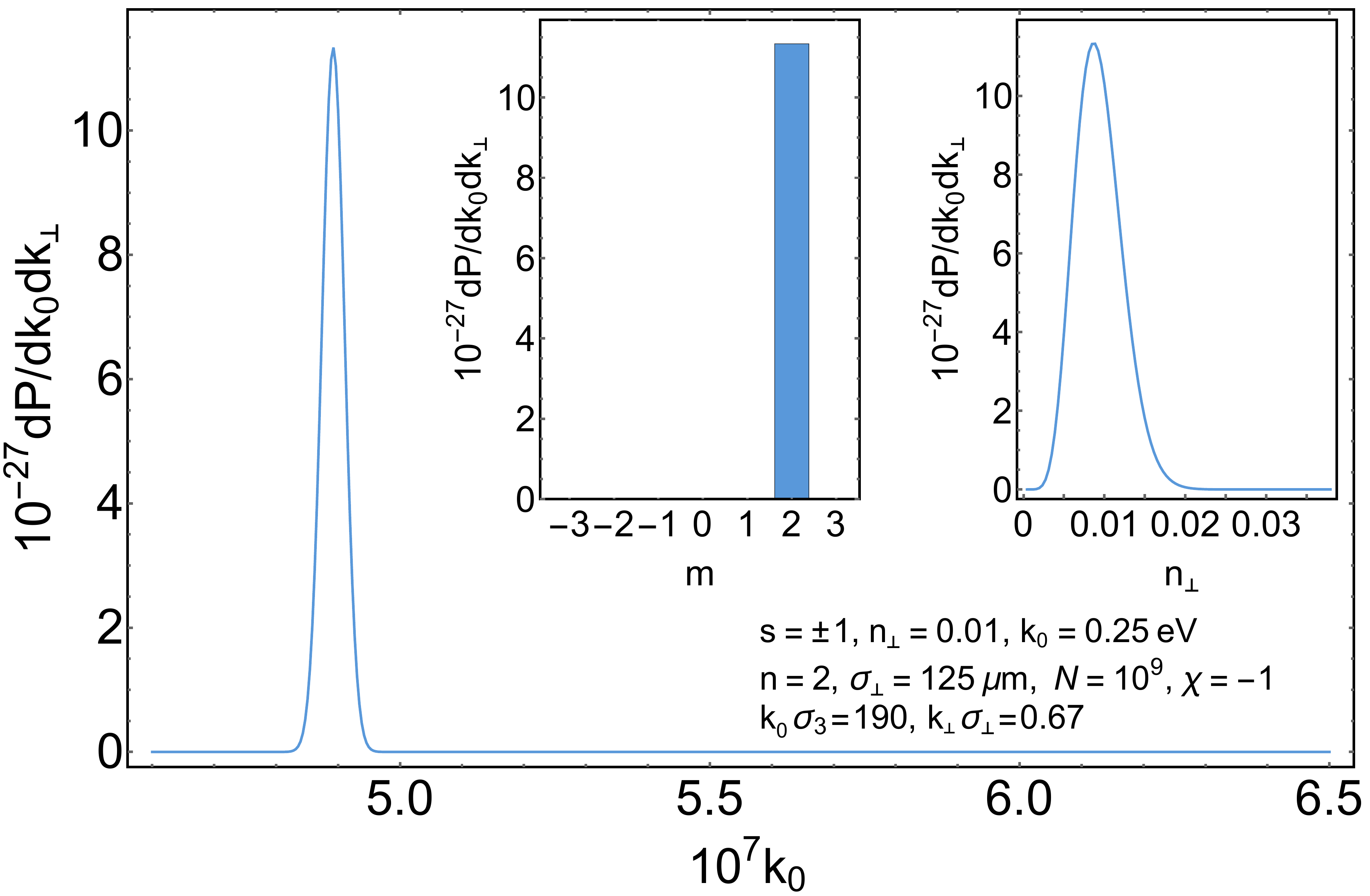}\;
\includegraphics*[align=c,width=0.475\linewidth]{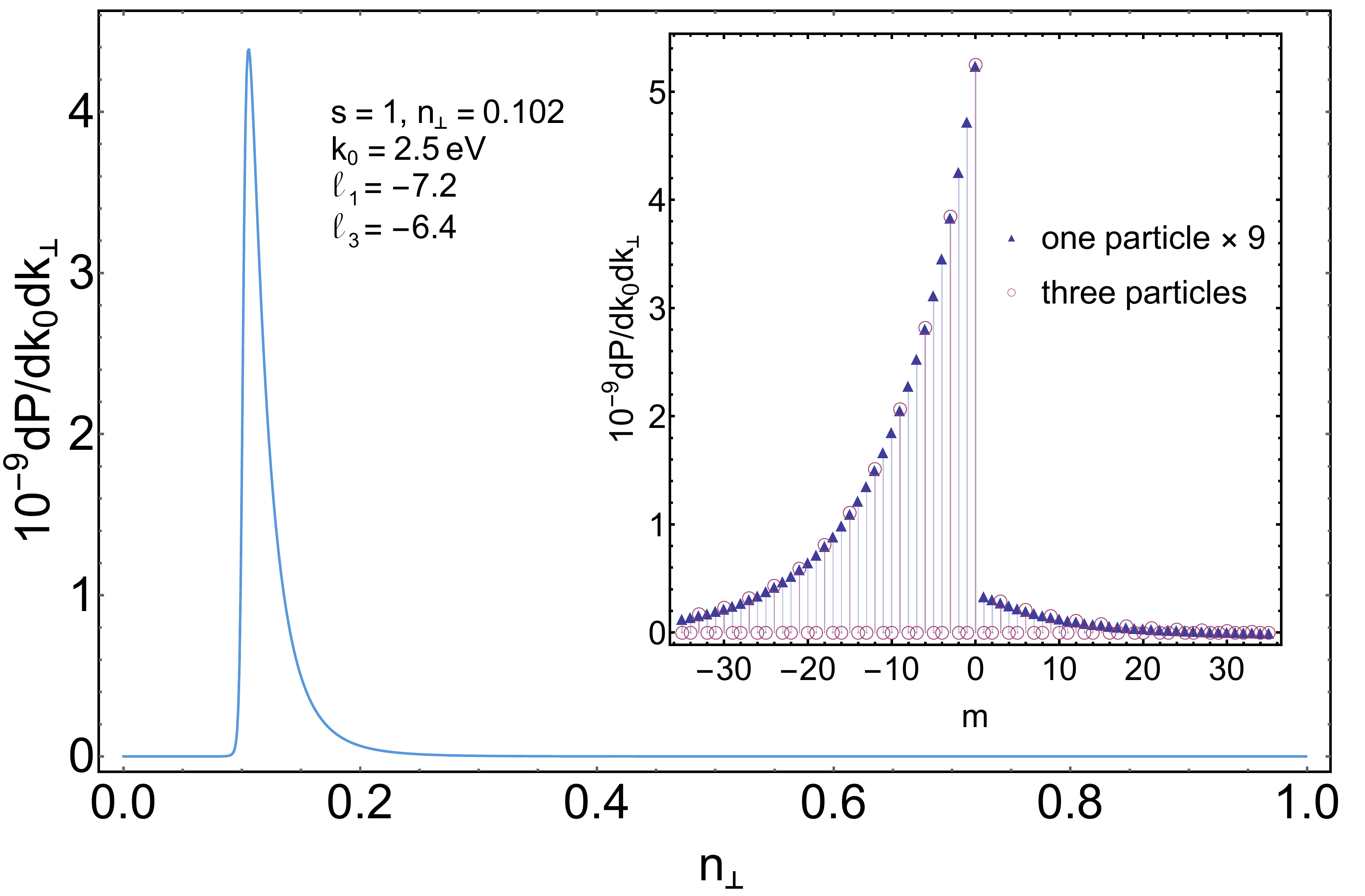}
\caption{{\footnotesize The transition radiation produced by charged particles striking a conducting plate. Left panel: The transition radiation of a helically microbunched beam of particles falling normally onto the conducting plate. The parameters of the particle beam are the same as in Fig. \ref{diel_plate2_fig}. The strong addition rule is evidently satisfied. The radiation probability is independent of the photon helicity $s$. In the paraxial regime, $n_\perp\ll1$, which obviously holds in our case, the orbital angular momentum can be introduced as $l=m-s$. Thus the radiation summed over the photon helicities is an equiprobable mixture of twisted photons with $l=\{1,3\}$. The twisted photons with definite value of the orbital angular momentum $l$ can be extracted from this radiation with the help of a circular polarizer selecting $s=1$ or $s=-1$. Right panel: The transition radiation of three charged particles falling onto the conducting plate at the angle $\theta=1/10$. The trajectories of these particles are obtained one from another by rotation around the detector axis by an angle of $2\pi/3$. The fulfilment of the selection rules discussed in \cite{BKL3} is shown. The maximum of the projection of the total angular momentum per photon is reached at $n_\perp\approx K(1+1/(\sqrt{2}K))/\ga\approx0.102$, where $K:=\ga\beta\sin\theta$.}}
\label{cond_plate_fig}
\end{figure}

The probability to record a twisted photon produced by a charged particle moving along the trajectory \eqref{part_traj} is given by \eqref{prob_plate}. The expression under the modulus sign in \eqref{prob_plate} becomes
\begin{equation}\label{ampl_cond}
    A:=-\frac{i|\be_3|}{k_0}\Big[\frac{1}{1-n_3\be_3} +\frac{1}{1+n_3\be_3}\Big]\de_{0,m}=-\frac{i|\be_3|}{k_0}\frac{2}{1-(n_3\be_3)^2}\de_{0,m}.
\end{equation}
This expression has a clear physical interpretation (see, e.g., \cite{GinzbThPhAstr,GinzTsyt}): the two terms in \eqref{ampl_cond} correspond to the contributions of the edge radiation produced by the charge and its image with an opposite charge in the ideal conductor (the mirror). The probability to record the twisted photon is
\begin{equation}
    dP(s,m,k_\perp,k_3)= \frac{4e^2\be^2_3\de_{0,m}}{k_0^2(1-(n_3\be_3)^2)^2}\Big(\frac{k_\perp}{2k_0}\Big)^3\frac{dk_3 dk_\perp}{2\pi^2}.
\end{equation}
It is concentrated at $m=0$ and does not depend on the photon helicity. Just as in the case of a dielectric, the probability of radiation of twisted photons by a bunch of charged particles moving along parallel trajectories normal to the surface of the ideal conductor has the form \eqref{prob_rad_tot_hw}. The use of helically microbunched beams of particles allows one to shift the coherent radiation probability distribution over $m$ by the signed number of coherent harmonic \cite{HemRos09,BKLb} (see Fig. \ref{cond_plate_fig}). The sum rule \eqref{ang_mom_per_phot} with $\ell_1=0$ is fulfilled.

If the charged particle moves at an angle to the detector axis along the trajectory \eqref{traj_tilt}, then the expression under the modulus sign in \eqref{prob_plate} is reduced to
\begin{equation}
    A=I(\be_{3})-I(-\be_{3})=\sgn(\be_3)\frac{i^{-1-m} e^{i m \de}}{k_{\perp}n_{\perp}}\Big[\Big(\frac{\be_{3}-n_{3}}{\kappa(n_{3},\be_{3})}-s \sgn(m)\Big)  q^{|m|}(n_{3},\be_{3})-(\be_{3}\leftrightarrow -\be_{3})\Big],
\end{equation}
The first term comes from the edge radiation of the charge and the second one is from its image in the mirror. Then the probability to record a twisted photon is given by \eqref{probab_plate_thck} with the normalization factor $|a|^2=1$. The reflection symmetry \eqref{refl_symm} holds for the radiation created by an arbitrary number of charged particles moving along the trajectories \eqref{traj_tilt} with different velocities $\boldsymbol{\be}$.

It was pointed out in \cite{BKL3} that the edge radiation can be used as a superradiant source of soft twisted photons. As we see, the use of a conducting mirror allows one to simplify the configuration of particle beams generating twisted photons in the spectral range where the conductor can be regarded as ideal. For example, if one focuses the three beams of charged particles in one point on the mirror such that the angles of incidence are equal and the angles between the beams are the same, then the probability distribution over $m$ of recorded twisted photons is asymmetric for a given $s$ (see Fig. \ref{cond_plate_fig}), obeys the selection rule $m=3k$, $k\in \mathbb{Z}$, and the reflection symmetry \eqref{refl_symm}. The absolute value of projection of the total angular momentum per photon is of order $K/(2\sqrt{2})$ \cite{BKL3}, where $K:=\ga\be_\perp$ and $\be_\perp=|\be_+|$.

As for an ideal conductor with the surface of a general form, the solution of the problem \eqref{Max_eqns}, \eqref{bound_conds_cond} can be reduced to the solution of the Maxwell equations \eqref{Max_eqns} with a singular source. The nontrivial boundary conditions on the closed hypersurface $\Si$ are standardly replaced by the singular source
\begin{equation}\label{sing_sourc}
    j^i[\psi;\spx)=\int_\Si d\s\sqrt{h}\Big\{n^{[i}\pr^{j]k}\partial_j\de(\spx-\spx(\s))\psi_k +\de(\spx-\spx(\s))\de^{ik}\partial_{[k}\psi_{j]}n^j \Big\},
\end{equation}
where $x^i(\s)$, $\s=\{\s^a\}$, $a=\overline{1,2}$, is the embedding map of the closed hypersurface $\Si$ into $\mathbb{R}^3$, $n^i$ is the unit normal to $\Si$ directed into the conductor,
\begin{equation}
    \pr^{ij}:=h^{ab}\partial_ax^i\partial_bx^j,
\end{equation}
where $h^{ab}$ is the inverse to the induced metric $h_{ab}=\partial_ax^i\partial_bx^j\de_{ij}$, and $h=\det h_{ab}$. Taking into account the boundary conditions \eqref{bound_conds_cond}, we obtain
\begin{equation}
    j_i[\psi;\spx)=\int_\Si d\s\sqrt{h}n^j\de(\spx-\spx(\s))\partial_{[i}\psi_{j]}.
\end{equation}
From the physical point of view, this current describes the currents induced by the external electromagnetic field on the ideal conductor. The corresponding vacuum Maxwell equations become
\begin{equation}
    (k_0^2-\hat{h}^2)\psi_i(k_0,\spx)-j_i[\psi;\spx)=0.
\end{equation}
The resolvent (the Green function) for this equation can be found perturbatively regarding the contribution $j_i[\psi]$ as a perturbation.

\subsection{Helical medium}\label{Hel_Medium}

A promising pure source of twisted photons with large projection of the total angular momentum $m$ is the radiation produced by charged particles in helical media. Let the permittivity $\e(k_0,\spx)$ be invariant under the transformations
\begin{equation}\label{spiral_trans}
    z\rightarrow z+\vf_r/q_0=z+\frac{\la_0}{r},\qquad\vf\rightarrow\vf+\vf_r,
\end{equation}
where $q_0=2\pi/\la_0$, $\vf$ is the azimuth angle of the cylindrical system of coordinates with the reference axis $z$, and $\vf_r=2\pi/r$, $r=\overline{1,\infty}$, is a fixed rotation angle. We will call the medium possessing such a symmetry as helical medium. We denote as $V_r$ the unitary operator acting in the space of solutions to the Maxwell equations and realizing the symmetry transformation \eqref{spiral_trans}. It follows from \eqref{spiral_trans} that $\e(k_0,\spx)$ is a periodic function of $z$ with the period $|\la_0|$. Of course, such a situation is never realized. However, if the number of periods $N\gtrsim10$, then, in describing the radiation generated by charged particles, the edge effects can be neglected (see below).

In accordance with the standard theorems (see, e.g., \cite{LandLifshQM.11}), the complete set of solutions of the Maxwell equations \eqref{Max_eqns} or \eqref{Max_eqns1} can be found in the form of the eigenfunctions of the symmetry operator
\begin{equation}\label{eigen_probl_V}
     \hat{V}_r\psi=e^{i\xi}\psi.
\end{equation}
We shall assume that the complex vector field $\psi$ is periodic with respect to the variable $z$ with the period $N|\la_0|$ for sufficiently large $N$. Developing the mode functions as a Fourier series
\begin{equation}
    \psi_3(\rho,\vf,z)=\sum_{k,n}a_{kn}(\rho)e^{ik\vf}e^{in\frac{q_0}{N}z},
\end{equation}
where $\rho$ is the distance from the $z$ axis, and substituting into \eqref{eigen_probl_V}, we obtain
\begin{equation}
    \xi=\frac{2\pi}{Nr}\vk,\quad\vk\in \mathbb{Z},
\end{equation}
and
\begin{equation}
    \psi_3(\rho,\vf,z)=\sum_{k,n}\tilde{a}_{k,N(rn-k)+\vk}(\rho)e^{ik(\vf-q_0z)}e^{i(\frac{\vk}{N} +rn )q_0z}.
\end{equation}
Introducing
\begin{equation}
    k_3=\frac{q_0}{N}\vk,
\end{equation}
and shifting $k\rightarrow k+m$, $\vk\rightarrow\vk+mN$, where $m\in \mathbb{Z}$, we have
\begin{equation}\label{mode_fourier}
    \psi_3(m,k_3;\rho,\vf,z)=\sum_{k,n}\bar{a}_{k+m}\big(k_3+q_0(rn-k);\rho\big) e^{ik(\vf-q_0z)}e^{im\vf}e^{i(k_3 +q_0rn )z}.
\end{equation}
As a result,
\begin{equation}\label{f3}
    \psi_3(m,k_3;\rho,\vf,z)=f_3(m,k_3;\rho,z,\vf-q_0z)e^{ik_3z}e^{im\vf},
\end{equation}
where
\begin{equation}\label{periodicity}
    f_3(m,k_3;\rho,z,\vf)=f_3\Big(m,k_3;\rho,z+\frac{\la_0}{r},\vf\Big) =f_3(m,k_3;\rho,z,\vf+2\pi).
\end{equation}
The same procedure applied to the transverse components of the mode functions leads to
\begin{equation}
    \psi_\pm(m,k_3;\rho,\vf,z)=f_\pm(m,k_3;\rho,z,\vf-q_0z)e^{ik_3z}e^{i(m\pm1)\vf},
\end{equation}
where the functions $f_\pm$ obey the periodicity conditions \eqref{periodicity}.

Further, we suppose that the permittivity $\e(k_0,\spx)$ tends sufficiently fast to unity as $\rho\rightarrow+\infty$. The probability to record a twisted photon by the detector is determined only by the scattering states, which tend to the solutions of the free Maxwell equations as $\rho\rightarrow+\infty$. For these states, we have
\begin{equation}\label{asymptotics}
    f_3(m,k_3;\rho,z,\vf)\rightarrow f_3(m,k_3;\rho),\qquad f_\pm(m,k_3;\rho,z,\vf)\rightarrow f_\pm(m,k_3;\rho),
\end{equation}
as $\rho\rightarrow+\infty$, i.e., the dependence of functions $f_{3,\pm}$ on the last two arguments disappears in this limit. It is this asymptotics that conditions the presence of factors
\begin{equation}
    e^{ik_3z}e^{im\vf},\qquad e^{ik_3z}e^{i(m\pm1)\vf}
\end{equation}
in $\psi_3$ and $\psi_\pm$, respectively.

If, additionally, the permittivity $\e(k_0,\spx)$ is symmetric under the rotations
\begin{equation}\label{rot_trans}
    \vf\rightarrow\vf+\vf_q,
\end{equation}
where $q$ is some natural number, then expansion \eqref{mode_fourier} turns into
\begin{equation}\label{mode_fourier_1}
    \psi_3(m,k_3;\rho,\vf,z)=\sum_{k,n}b_{qk+m}\big(k_3+q_0(rn-qk);\rho\big) e^{iqk(\vf-q_0z)}e^{im\vf}e^{i(k_3 +q_0rn )z},
\end{equation}
formula \eqref{f3} is left intact, and the periodicity property \eqref{periodicity} becomes
\begin{equation}
    f_3(m,k_3;\rho,z,\vf)=f_3\Big(m,k_3;\rho,z+\frac{\la_0}{r},\vf\Big) =f_3(m,k_3;\rho,z,\vf+2\pi/q).
\end{equation}
The analogous formulas are valid for the transverse components $\psi_\pm$.

Now we can use general formula \eqref{probab_point}. As long as the integral entering into this formula is saturated in the region $\Omega$ where the asymptotics \eqref{asymptotics} hold, the amplitude of radiation of a twisted photon with the momentum $k_3$ and the projection of the total angular momentum $m$ by a charged particle moving along the trajectory \eqref{part_traj} is proportional to
\begin{equation}\label{helic_rad}
    \int_\infty^\infty dt e^{-ik_0 t}\psi_3(m,k_3;0,\vf,\be_3 t).
\end{equation}
Substituting expansion \eqref{mode_fourier_1} into this integral and bearing in mind that the dependence of the mode functions on the azimuth angle $\vf$ must disappear at $\rho=0$, we conclude that integral \eqref{helic_rad} is proportional to
\begin{equation}
    \de\big(k_0-\big(k_3+q_0(m+rn)\big)\be_3 \big),\qquad m=-qk.
\end{equation}
To put it differently, the radiation of twisted photons is concentrated at the harmonics
\begin{equation}\label{sel_rul_1}
    k_0=|q_0|\frac{|\be_3| \bar{n}}{1-\be_3n_3}=\frac{2\pi}{|\la_0|}\frac{|\be_3| \bar{n}}{1-\be_3n_3},
\end{equation}
where $\bar{n}=\overline{1,\infty}$ is the harmonic number, and the following selection rule holds:
\begin{equation}\label{sel_rul_2}
    m=\sgn(\la_0\be_3)\bar{n}+r n=q k,
\end{equation}
where $n$ and $k$ are some integer numbers. The intensity of radiation at these harmonics is determined by the form of the permittivity $\e(k_0,\spx)$.

The case of the permittivity invariant with respect to the continuous transformations
\begin{equation}\label{spiral_trans_1}
    z\rightarrow z+\psi/q_0,\qquad\vf\rightarrow\vf+\psi,
\end{equation}
where $\psi\in \mathbb{R}$, is formally obtained from \eqref{spiral_trans} in the limit $r\rightarrow\infty$. Then the selection rule \eqref{sel_rul_1} is left unchanged and \eqref{sel_rul_2} is replaced by
\begin{equation}\label{sel_rul_cont}
    m=\sgn(\la_0\be_3)\bar{n}=q k.
\end{equation}
As we see, the radiation at the $\bar{n}$-th harmonic is a pure source of twisted photons with a definite value of the projection of the total angular momentum $m$.

\begin{figure}[tp]
\centering
\includegraphics*[align=c,width=0.2\linewidth]{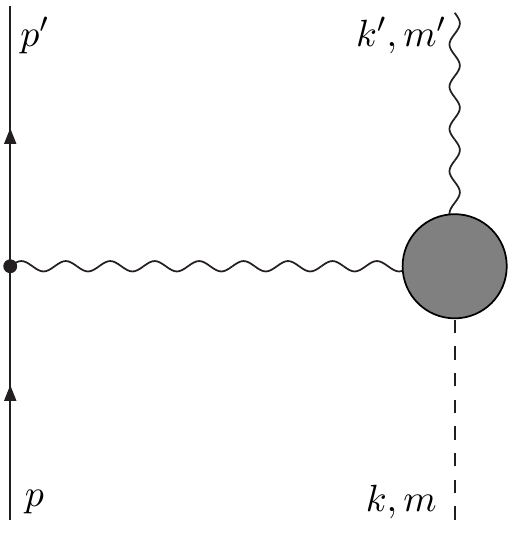}\qquad\qquad
\includegraphics*[align=c,width=0.2\linewidth]{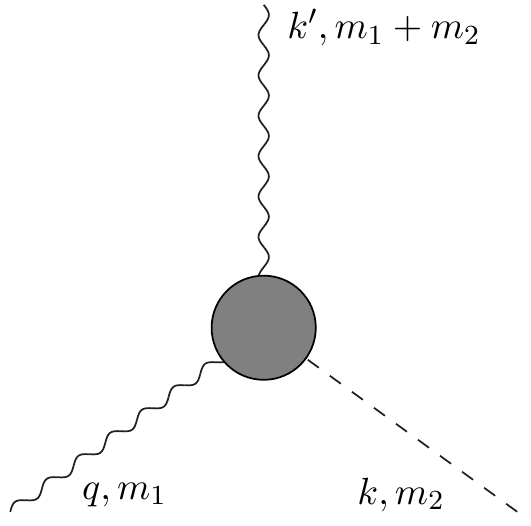}
\caption{{\footnotesize Left panel: The Feynman diagram for the leading contribution to transition scattering of the permittivity wave on the charged particle. The permittivity wave is denoted as a dashed line and the blob denotes the interaction vertex of the permittivity with the electromagnetic field. This vertex is the first nontrivial term of the expansion of the photon propagator in a medium in powers of susceptibility. Right panel: The first order diagram describing the conversion of a plane-wave photon to a twisted one. For a normal incidence of the plane-wave photon onto the medium interface, the projection of the total angular momentum $m_1=s_1$, where $s_1$ is the photon helicity. This process was employed in \cite{Barboza15,Baboza12,YLiu18} for generation of twisted photons.}}
\label{diag_1}
\end{figure}

The above derivation is completely applicable to the case when the permittivity is not isotropic and the permittivity tensor has the form
\begin{equation}\label{permit_holec}
    \e_{ij}(k_0,\spx)=\e_\perp(k_0,\spx)\de_{ij}+[\e_\parallel(k_0,\spx)-\e_\perp(k_0,\spx)]n_in_j,
\end{equation}
where the functions $\e_\perp(k_0,\spx)$, $\e_\parallel(k_0,\spx)$ are invariant under \eqref{spiral_trans_1} and
\begin{equation}
    n_i=\big(\cos(q_0 z),\sin(q_0 z),0\big).
\end{equation}
The permittivity tensor \eqref{permit_holec} is inherent to cholesteric liquid crystals (see, e.g., \cite{deGenProst}). Therefore, the radiation produced by a charged particle moving along a helical axis of the cholesteric is a pure source of twisted photons obeying the selection rules \eqref{sel_rul_1}, \eqref{sel_rul_cont} with $q=1$ and $|\la_0|$ being equal to a half of the chiral pitch. Notice that the cholesteric liquid crystals were used in \cite{Barboza15,Baboza12,VoloLavr} to convert an ordinary laser wave to a twisted one. In fact, we consider the process depicted by the left diagram in Fig. \ref{diag_1}, while the conversion of plane-wave photons is described by the right diagram in Fig. \ref{diag_1}.

Another realization of a medium possessing symmetry \eqref{spiral_trans_1} is represented by the conductor in the form of a helix with $q$ branches. The bunch of charged particles move along the axis of this helix. The conductors of such a form were used in \cite{YLiu18} to transform ordinary plane-wave photons to twisted ones. The medium with permittivity symmetric with respect to \eqref{spiral_trans} can also be made by arranging properly the dielectrics with different $\e$. One more means to construct a helical medium with symmetry \eqref{spiral_trans_1} is to deform the medium in a twisted manner by using, for example, the helical sound waves. The wave packet of twisted phonons, which is sufficiently wide in $m$ and obeys \eqref{eigen_probl_V} with good accuracy, represents a helical sound wave and leads to an appropriate variation of permittivity. As a rule, the phonon velocity can be ignored in comparison with the velocity of a relativistic particle and so the formulas derived above are applicable to this case without any modifications. Thereby the twisted photons are generated by transition scattering \cite{GinzbThPhAstr,GinzTsyt} of helical waves of permittivity on charged particles (see Fig. \ref{diag_1}). The same situation happens in a plasma perturbed by the helical wave of phonons or photons \cite{Pisanty19,CLiu16,XLZhu18,LBJu16}. However, in order to describe the radiation of twisted photons quantitatively, one needs to take a spatial dispersion into account in this case.

The selection rules \eqref{sel_rul_1}, \eqref{sel_rul_2} coincide exactly with the selection rules found in \cite{BKL3} for the scattering of particles on helical targets. It is not surprising as \eqref{sel_rul_1}, \eqref{sel_rul_2} can be deduced from the following considerations (a similar derivation of selection rules for the radiation produced by charged particles in a layered medium can be found, e.g., in \cite{BazylZhev,TMikaelian}). Let the charged particle move in a medium along the $z$ axis with constant velocity, the $z$ axis coinciding with the detector axis. We assume that the medium permittivity is invariant under transformations \eqref{spiral_trans}, \eqref{rot_trans}. The Coulomb field of the charged particle induces the current in the medium. In virtue of the symmetry of the problem, the density of the total current $j^i$, which is the sum of the current density of the charged particle and the density of the induced current, is symmetric with respect to the transformations
\begin{equation}
    z\rightarrow z+\frac{\la_0}{2\pi}\vf_r,\qquad\vf\rightarrow\vf+\vf_r,\qquad x^0\rightarrow x^0+\frac{\la_0}{2\pi\be_3}\vf_r.
\end{equation}
The components of the current density transform as
\begin{equation}\label{cur_symm}
    j_3\rightarrow j_3,\qquad j_\pm\rightarrow j_\pm e^{\pm i\vf_r}.
\end{equation}
The mode functions of the vacuum twisted photons \eqref{tw_phot_vac} change accordingly
\begin{equation}\label{mode_symm}
\begin{split}
    \psi_3(m,k_3,k_\perp)&\rightarrow\psi_3(m,k_3,k_\perp)e^{-i(k_0\frac{\la_0}{2\pi\be_3}-k_3\frac{\la_0}{2\pi}-m)\vf_r},\\ \psi_\pm(s,m,k_3,k_\perp)&\rightarrow\psi_\pm(s,m,k_3,k_\perp)e^{-i(k_0\frac{\la_0}{2\pi\be_3}-k_3\frac{\la_0}{2\pi}-m)\vf_r}e^{\pm i\vf_r}.
\end{split}
\end{equation}
These relations allow us to rewrite the probability to record a twisted photon produced by the current $j^i$ in such a way that the fulfillment of the selection rules \eqref{sel_rul_1}, \eqref{sel_rul_2} becomes evident.

Indeed, the probability to record a twisted photon is given by the general formula
\begin{multline}\label{rad_prob}
    dP(s,m,k_3,k_\perp)=\bigg|\int d^4x e^{-ik_0x^0+ik_3x_3}\times\\
    \times\Big[\frac12 a_+(s,m,k_3,k_\perp;\spx)j_-(x) +\frac12a_-(s,m,k_3,k_\perp;\spx)j_+(x)+a_3(m,k_3,k_\perp;\spx)j_3(x)\Big]\bigg|^2\Big(\frac{k_\perp}{2k_0}\Big)^3 \frac{dk_3dk_\perp}{2\pi^2},
\end{multline}
where the renormalized mode functions $a_3$, $a_\pm$ have been introduced (see (35) of \cite{BKL2}). Then we consider the integration over $N$ periods in \eqref{rad_prob} and neglect the edge effects. Parting the integral over $x^0$ as
\begin{equation}
    \int_0^{N|\la_0|} dx^0\cdots=\sum_{n=1}^{Nr} \int_{|\la_0|(n-1)/(r|\be_3|)}^{|\la_0|n/(r|\be_3|)} dx^0\cdots=:\sum_{n=1}^{Nr} A_n,
\end{equation}
and employing the symmetry properties \eqref{cur_symm}, \eqref{mode_symm}, we find that
\begin{equation}
    A_n=e^{-i(k_0\frac{\la_0}{2\pi\be_3}-k_3\frac{\la_0}{2\pi}-m)\sgn(\la_0\be_3)\vf_r(n-1)} A_1,
\end{equation}
where
\begin{multline}
    A_1=\int_0^{|\la_0|/(r|\be_3|)}dx^0\int d\spx e^{-ik_0x^0+ik_3x_3}\times\\
    \times\Big[\frac12 a_+(s,m,k_3,k_\perp;\spx)j_-(x) +\frac12a_-(s,m,k_3,k_\perp;\spx)j_+(x)+a_3(m,k_3,k_\perp;\spx)j_3(x)\Big].
\end{multline}
As a result, the radiation probability \eqref{rad_prob} is written as
\begin{equation}
    dP(s,m,k_3,k_\perp)=\sum_{n=1}^{Nr}\Big|e^{-i(\frac{k_0}{q_0\be_3}-\frac{k_3}{q_0}-m)\sgn(\la_0\be_3)\vf_r(n-1)}\Big|^2 dP_1(s,m,k_3,k_\perp),
\end{equation}
where
\begin{equation}
    dP_1(s,m,k_3,k_\perp):=|A_1|^2\Big(\frac{k_\perp}{2k_0}\Big)^3 \frac{dk_3dk_\perp}{2\pi^2}.
\end{equation}
The interference factor
\begin{equation}
\begin{split}
    I(m,k_3,k_\perp)&=\sum_{n=1}^{Nr}\Big|e^{-i(\frac{k_0}{q_0\be_3}-\frac{k_3}{q_0}-m)\sgn(\la_0\be_3)\vf_r(n-1)}\Big|^2 =\frac{\sin^2(\pi N\de)}{\sin^2(\pi\de/r)},\\
    \de:&=m-\frac{k_0\la_0}{2\pi}(\be_3^{-1}-n_3),
\end{split}
\end{equation}
has the same form as in \cite{BKL3}. It possesses the sharp maxima at $\de=nr$, where $n\in \mathbb{Z}$. Taking into account that  permittivity is invariant under rotations \eqref{rot_trans}, the condition $\de=nr$ leads to the selection rules \eqref{sel_rul_1}, \eqref{sel_rul_2}. At the maxima, the interference factor is equal to
\begin{equation}
    I_{max}=N^2r^2,
\end{equation}
whence it is evident that, for sufficiently large $N$, the edge effects give a negligible contribution to radiation at harmonics \eqref{sel_rul_1}.

The selection rules \eqref{sel_rul_1}, \eqref{sel_rul_2} can also be deduced with the aid of the conservation laws of the momentum and total angular momentum applied to the process of transition scattering of the wave of permittivity on the charged particle moving uniformly along a straight line \cite{GinzTsyt}. Let us consider a stationary permittivity tensor invariant with respect to \eqref{spiral_trans}. Then the Fourier modes of the wave corresponding to such a permittivity tensor obey the relations
\begin{equation}\label{pump}
    k_0=0,\qquad (m+k_3/ q_0)\vf_r=2\pi n\;\Rightarrow\;k_3=q_0(nr-m),\quad n\in \mathbb{Z},
\end{equation}
where $m$ is the projection of the total angular momentum onto the twisted photon detector axis. Hence, for the diagram depicted in Fig. \ref{diag_1}, which describes the leading order contribution to the twisted photon radiation by a charged particle moving along the detector axis, we have
\begin{equation}\label{cons_laws}
    M\ga \be_3+k_3=M\ga' \be'_3+k'_3,\qquad M\ga=M\ga'+k'_0,\qquad m=m',
\end{equation}
where $M$ is the charged particle mass, $\ga$ is its Lorentz factor, and the prime marks the quantities after scattering. The last equality in \eqref{cons_laws} expresses conservation of projection of the total angular momentum and we suppose that the charged particle does not change its projection of the total angular momentum. Assuming that $k'_0/(M\ga)\ll1$, we have from \eqref{cons_laws}:
\begin{equation}
    k_3=k'_3-k'_0/\be_3.
\end{equation}
Therefore,
\begin{equation}
    k'_0=q_0\be_3\frac{m-nr}{1-\be_3 n'_3},\qquad n'_3:=k'_3/k'_0.
\end{equation}
Denoting as
\begin{equation}
    \bar{n}:=\sgn(\la_0\be_3)(m-nr),
\end{equation}
we arrive at the selection rules \eqref{sel_rul_1}, \eqref{sel_rul_2} with $q=1$. The generalization to the case $q\in \mathbb{N}$ is obvious.

The above selection rules were obtained for radiation of twisted photons by one particle or by sufficiently narrow uniform beam of them in a helical medium. The use of sufficiently long helically microbunched beams allows one to shift the radiation probability distribution over $m$ in accordance with the strong addition rule \cite{BKLb}.

\section{Conclusion}

Let us sum up the results. In Secs. \ref{Quant_in_Medium} and \ref{Recording_Tw_Phot}, we developed a general quantum theory of radiation of twisted photons by charged particles propagating in an inhomogeneous dispersive medium. The production of twisted photons in homogeneous dispersive media was already studied in \cite{IvSerZay,Kaminer16}, but the inhomogeneity of a medium was not taken into account. As we have already discussed in the introduction, one usually needs a source of twisted photons in a vacuum and not in the medium. Therefore, it is relevant to consider the production of twisted photons in inhomogeneous media since the twisted photons can be destroyed by the inhomogeneities of permittivity.

In Sec. \ref{Recording_Tw_Phot}, we derived the general formula for probability to record a twisted photon produced by a classical current in an inhomogeneous dispersive medium. In the case of a homogeneous medium investigated in Sec. \ref{Homogeneous_Medium}, we reproduced the known results. In Sec. \ref{Examples}, we applied the general formula to description of radiation of twisted photons in several particular configurations.

In Secs. \ref{Diel_Plate} and \ref{Thick_Diel_Plate}, we deduced the explicit expressions for the probability to record a twisted photon produced by a charged particle moving with constant velocity and crossing a dielectric plate. The axis of the detector of twisted photons was assumed to be normal to the surface of the dielectric plate. As expected, we found that, in the case when a charged particle moves along the detector axis, all the radiated twisted photons possess a zero projection of the total angular momentum and the probability to record a twisted photon does not depend on the photon helicity. Moreover, we proved that the probability to record a twisted photon produced by an arbitrary number of charged particles moving uniformly along straight lines intersecting at one point of the surface of a thick transparent dielectric plate possesses the reflection symmetry \cite{BKL3}. Such a radiation describes an infrared asymptotics of radiation of twisted photons in an arbitrary QED process near a dielectric plate. Of course, for this asymptotics takes place, it is necessary that the wavelength of radiated photons will be much smaller than the typical sizes of the dielectric plate.

In Sec. \ref{Cond_Plate}, we considered the radiation of twisted photons by a charged particle falling onto an ideally conducting plate. In fact, the radiation in this case is the edge radiation completely described in terms of twisted photons in \cite{BKL3}. As in the case of the edge radiation, the probability to record a twisted photon obeys the reflection symmetry.

In Sec. \ref{Hel_Medium}, we investigated the radiation produced by charged particles moving along the axis of a twisted photon detector in a helical medium. A typical example of such a medium is a cholesteric liquid crystal (see, e.g., \cite{deGenProst}). Using different approaches, we proved that, in this case, the radiation of twisted photons obeys the same selection rules as were found in \cite{BKL3} for the scattering of charged particles on helical targets. We provided a simple explanation of this fact in terms of transition scattering on a helical permittivity wave. These specific properties of helical media can be employed for elaboration of the active medium for coherent amplification of stimulated radiation of twisted photons.

Employing the general formulas \cite{BKb,BKLb} for the radiation of twisted photons by particle beams, we also described the radiation produced by uniform Gaussian and helically microbunched beams of relativistic charged particles. In particular, we showed explicitly the fulfillment of the strong addition rule \cite{BKLb,RibGauNin14,HemMar12,HemMarRos11,HemRos09,HemStuXiZh14,HKDXMHR,HemsingTR12} for the radiation of twisted photons by helical beams of particles falling normally onto a medium surface. Such beams can be used to generate VCh and transition radiations with large projections of the total angular momentum.

\paragraph{Acknowledgments.}

The work was supported by the Ministry of Science and Higher Education of the Russian Federation, Project No. 3.9594.2017/8.9. The numerical simulations of radiation of twisted photons by beams of particles were performed on TPU Supercomputing Cluster within the framework of Tomsk Polytechnic University Competitiveness Enhancement Program.

\appendix
\section{Interference factors for Gaussian particle beams}\label{Inter_Fact_App}

The general formulas for the incoherent, $f_{mn}$, and coherent, $\vf_m$, interference factors read as \cite{BKLb}
\begin{equation}\label{interfer_factor}
\begin{split}
    f_{mn}&=\int d\spb\rho(\spb)j^*_m\big(k_\perp\de_+(\spb),k_\perp\de_-(\spb)\big) j_n\big(k_\perp\de_+(\spb),k_\perp\de_-(\spb)\big),\\
    \vf_m&=\int d\spb\rho(\spb) e^{ik_0\de^0(\spb)-ik_3\de^3(\spb)} j^*_m\big(k_\perp\de_+(\spb),k_\perp\de_-(\spb)\big).
\end{split}
\end{equation}
We particularize these formulas to the case when the normal $\boldsymbol{\xi}=(0,0,1)$ to the vacuum-medium interface at the points where the beam of particles enters and exits the medium is parallel to the axis of the twisted photon detector. The particle beam moves initially with the velocity
\begin{equation}
    \boldsymbol\be=\be(\sin\theta,0,\cos\theta).
\end{equation}
The one-particle probability distribution has the form (see Sec. 3.5 of \cite{BKLb})
\begin{equation}\label{particle_probab}
    \rho(\spb')=\frac{e^{-b'^2_3/(2\s_3^2)}}{\sqrt{2\pi}\s_3}\frac{e^{-b'_+b'_-/(2\s_\perp^2)}}{2\pi\s_\perp^2} \sideset{}{'}\sum_{k=0}^\infty \Big[\al_k \Big(\frac{b'_+}{\s_\perp}\Big)^ke^{2\pi i\chi kb'_3/\de} +\al_{-k} \Big(\frac{b'_-}{\s_\perp}\Big)^ke^{-2\pi i\chi kb'_3/\de}\Big],
\end{equation}
where $\al^*_k=\al_{-k}$ are given in (74) of \cite{BKLb}, the prime at the sum sign says that the term with $k=0$ should be multiplied by $1/2$, and
\begin{equation}
    b'_1=\cos\theta b_1-\sin\theta b_3,\qquad b'_2=b_2, \qquad b'_3=\sin\theta b_1+\cos\theta b_3.
\end{equation}
Using general formula (12) of \cite{BKLb}, we find
\begin{equation}
    \de_1=\frac{b'_1}{\cos\theta},\qquad \de_2=b'_2,\qquad \de_3=0,\qquad\de^0=\frac{\tg\theta b'_1 -b'_3}{\be}.
\end{equation}
The probability distribution \eqref{particle_probab} describes a helically microbunched particle beam with the helix pitch $\de$ and the handedness $\chi$. The case $\de\rightarrow\infty$ corresponds to a uniform Gaussian beam of particles.

It is convenient to change the integration variables in \eqref{interfer_factor} and pass from $\spb$ to $\spb'$. The Jacobian of this change equals unity. Having performed such a change of variables, we will not write the primes at $\spb'$ to shorten the notation. Let us start with the incoherent interference factor. First, we write
\begin{equation}
    j^*_m j_n=i^{m-n}\int_{-\pi}^\pi\frac{d\psi_1d\psi_2}{(2\pi)^2} e^{im\psi_1-in\psi_2} e^{ik_\perp b_2(\sin\psi_1-\sin\psi_2)}e^{ik_\perp b_1(\cos\psi_1-\cos\psi_2)/\cos\theta}.
\end{equation}
Due to periodicity of the integrand, the change of variables
\begin{equation}
    \frac{\psi_1-\psi_2}{2}\rightarrow\psi_1,\qquad \frac{\psi_1+\psi_2}{2}\rightarrow\psi_2,
\end{equation}
results in
\begin{equation}
    \int_{-\pi}^\pi\frac{d\psi_1d\psi_2}{(2\pi)^2}\rightarrow \int_{-\pi}^\pi\frac{d\psi_1d\psi_2}{(2\pi)^2}.
\end{equation}
Hence,
\begin{equation}
    j^*_m j_n=i^{m-n}\int_{-\pi}^\pi\frac{d\psi_1d\psi_2}{(2\pi)^2} e^{i(n-m)\psi_2+i(n+m)\psi_1} e^{-ik_\perp\sin\psi_1 (b_+z^*+b_-z)},
\end{equation}
where
\begin{equation}
    z:=\frac{\sin\psi_2}{\cos\theta}-i\cos\psi_2.
\end{equation}
The integrals over $b_\pm$ in \eqref{interfer_factor} become Gaussian and can be evaluated with the aid of the relations
\begin{equation}\label{compl_Gaussian_int}
\begin{split}
    \int\frac{d b_+ db_-}{2i}b_+^m e^{-b_+b_-/2 +\be_+b_-+\be_-b_+}&=2\pi(2\be_+)^m e^{2\be_+\be_-},\\
    \int\frac{db_+ db_-}{2i}b_-^m e^{-b_+b_-/2 +\be_+b_-+\be_-b_+}&=2\pi(2\be_-)^m e^{2\be_+\be_-},
\end{split}
\end{equation}
where $d b_+ db_-=2idb_1db_2$. Besides,
\begin{equation}
    \int \frac{db_3}{\sqrt{2\pi}\s_3}e^{-b_3^2/(2\s_3^2) \pm2\pi i\chi kb_3/\de}=e^{-2\pi^2k^2\s_3^2/\de^2}.
\end{equation}
Using these formulas, we find that the term at $\al_k$ is given by the integral
\begin{equation}
    i^{m-n}e^{-2\pi^2k^2\s_3^2/\de^2} \int_{-\pi}^\pi\frac{d\psi_1d\psi_2}{(2\pi)^2} e^{i(n-m)\psi_2+i(n+m)\psi_1}  (-2ik_\perp\sigma_\perp\sin\psi_1 z)^k e^{-2k^2_\perp\sigma^2_\perp\sin^2\psi_1 |z|^2}.
\end{equation}
The term at $\al_{-k}$ becomes
\begin{equation}
    i^{m-n}e^{-2\pi^2k^2\s_3^2/\de^2} \int_{-\pi}^\pi\frac{d\psi_1d\psi_2}{(2\pi)^2} e^{i(n-m)\psi_2+i(n+m)\psi_1}  (2ik_\perp\sigma_\perp\sin\psi_1 z^*)^k e^{-2k^2_\perp\sigma^2_\perp\sin^2\psi_1 |z|^2}.
\end{equation}
One integration in this double integral can be performed. The integral over $\psi_1$ is reduced to
\begin{equation}
    \int_{-\pi}^\pi \frac{d\psi}{2\pi}e^{i(n+m)\psi}\sin^k\psi e^{-a^2\sin^2\psi}=\Big(\frac{i}{2}\Big)^k
    \left\{
      \begin{array}{ll}
        \sum_{s=0}^k\frac{(-1)^sk!}{s!(k-s)!} e^{-a^2/2}I_{s+(n+m-k)/2}(\frac{a^2}{2}), & \frac{n+m-k}{2}\in \mathbb{Z}; \\[0.8em]
        0, & \frac{n+m-k}{2}\not\in \mathbb{Z}.
      \end{array}
    \right.
\end{equation}
As a result, we come to
\begin{multline}\label{incoh_inter_f}
    f_{mn}=i^{m-n}\int_{-\pi}^\pi \frac{d\psi}{2\pi} e^{i(n-m)\psi} \sideset{}{'}\sum_{k=0}^\infty (k_\perp\s_\perp)^k e^{-2\pi^2k^2\s_3^2/\de^2}[\al_k z^k + \al_{-k} z^{*k} ]\times\\
     \times\sum_{s=0}^k \frac{(-1)^sk!}{s!(k-s)!} e^{-k_\perp^2\s_\perp^2|z|^2}I_{s+(n+m-k)/2}(k_\perp^2\s_\perp^2|z|^2),
\end{multline}
where $I_k(x)$ is the modified Bessel function of the first kind and $\psi_2$ in the definition of $z$ should be replaced by $\psi$. The prime at the sum sign reminds us that the term at $k=0$ should be taken with the factor $1/2$ and the terms such that $n+m-k$ is an odd number must be omitted.

Notice that when
\begin{equation}\label{long_bunch}
    \pi\s_3/\de\gtrsim1,
\end{equation}
the terms with $k\neq0$ are strongly suppressed, i.e., the incoherent radiation is the same as for a round particle beam \cite{BKb} and
\begin{equation}
    f_{mn}\approx i^{m-n}\int_{-\pi}^\pi \frac{d\psi}{2\pi} e^{i(n-m)\psi}
    \left\{
      \begin{array}{ll}
        e^{-k_\perp^2\s_\perp^2|z|^2}I_{(n+m)/2}(k_\perp^2\s_\perp^2|z|^2), & \frac{n+m}{2}\in \mathbb{Z}; \\[0.8em]
        0, & \frac{n+m}{2}\not\in \mathbb{Z}.
      \end{array}
    \right.
\end{equation}
For $\theta\ll1$ or $\pi-\theta\ll1$, the incoherent interference factor \eqref{incoh_inter_f} reduces to the corresponding factor found in \cite{BKLb} in the case of a forward radiation.

Now we turn to the coherent interference factor. Upon substitution,
\begin{equation}
\begin{split}
     e^{ik_0\de^0(\spb)-ik_3\de^3(\spb)} j^*_m\big(k_\perp\de_+(\spb),k_\perp\de_-(\spb)\big)&=i^m\int_{-\pi}^\pi \frac{d\psi}{2\pi} e^{-im\psi} e^{ik_0(\tg\theta b_1-b_3)/\be} e^{-ik_\perp(b_2\sin\psi+b_1\cos\psi/\cos\theta)}=\\
     &=e^{-ik_0b_3/\be}\int_{-\pi}^\pi \frac{d\psi}{2\pi} e^{-im\psi} e^{ik_\perp(b_+\tilde{z}^* +b_-\tilde{z})/2},
\end{split}
\end{equation}
where
\begin{equation}
    \tilde{z}:= \frac{\tg\theta}{\be n_\perp} +\frac{\sin\psi}{\cos\theta}-i\cos\psi,
\end{equation}
the integrals over $\spb$ in \eqref{interfer_factor} become Gaussian. Then we have
\begin{equation}
    \int \frac{db_3}{\sqrt{2\pi}\s_3}e^{-b_3^2/(2\s_3^2) \pm2\pi i\chi kb_3/\de -ik_0 b_3/\be}=e^{-\s_3^2(k_0/\be \mp2\pi\chi k/\de)^2/2}.
\end{equation}
The integrals over $b_\pm$ can be performed by the use of formulas \eqref{compl_Gaussian_int}. After a little algebra, we arrive at
\begin{equation}\label{coher_int_fact}
    \vf_m=\int_{-\pi}^\pi \frac{d\psi}{2\pi} e^{-im\psi}e^{-k_\perp^2\s_\perp^2|\tilde{z}|^2/2} \sideset{}{'}\sum_{k=0}^\infty (ik_\perp\s_\perp)^k  [\al_k \tilde{z}^k e^{-\s_3^2(k_0/\be-2\pi\chi k/\de)^2/2} + \al_{-k} \tilde{z}^{*k} e^{-\s_3^2(k_0/\be+2\pi\chi k/\de)^2/2}],
\end{equation}
where the prime at the sum sign says that the term with $k=0$ should be taken with the factor $1/2$. The integral over $\psi$ can be represented as the series of products of the modified Bessel functions of the first kind but this representation does not facilitate the evaluation of \eqref{coher_int_fact}. Therefore, we do not write it here.

Notice that when condition \eqref{long_bunch} is satisfied, the coherent interference factor and the probability of coherent radiation are concentrated near the coherent harmonics
\begin{equation}\label{coher_harmcs}
    k_0=2\pi\chi n\be/\de,\qquad \chi n>0,\quad n\in \mathbb{Z}.
\end{equation}
The radiation probability at these harmonics is proportional to $|\al_n|^2$. However, in general, the strong addition rule \cite{BKLb} is not fulfilled. Only when
\begin{equation}
    \De\theta\ll1,\qquad\frac{|\tg\theta|}{\be n_\perp}\ll1,\qquad k_\perp^2\s_\perp^2\frac{|\tg\theta|}{\be n_\perp}\ll1,
\end{equation}
where $\De\theta=\theta$ or $\De\theta=\pi-\theta$, does the strong addition rule hold. This case, of course, corresponds to the forward radiation.


\end{document}